\documentclass[conference]{IEEEtran}
\IEEEoverridecommandlockouts

\usepackage{cite}
\usepackage{amsmath,amssymb,amsfonts}
\usepackage{graphicx}
\usepackage{textcomp}
\usepackage{xcolor}
\usepackage{verbatim}
\usepackage{algorithm}
\usepackage{algpseudocode}
\usepackage{multirow}
\usepackage{tikz}
\usetikzlibrary{positioning,calc,fit,backgrounds,arrows.meta}

\begin{document}

\title{Joint Design of Doppler-Resilient Unimodular Discrete-Phase Waveforms and Receiving Filters for MIMO Radars}

\author{Junpeng Ma, Yuke Li, \IEEEmembership{Member, IEEE}, Junbo Wang, \IEEEmembership{Member, IEEE}, Yongxing Zhou, \IEEEmembership{Fellow, IEEE} 
\thanks{Manuscript received April 22nd, 2026;  revised 00-00-0000; accepted 00-00-0000.
Date of publication 00-00-0000; date of current version April 22nd, 2026.}
\thanks{This work was jointly supported by the Fundamental Research Funds for the
Central Universities (D5000250060, D5000250044), Natural Science Basic
Research Program of Shaanxi (2025JC-YBQN-805, 2025JC-YBQN-882).}
\thanks{Junpeng Ma, Yuke Li and Junbo Wang are with The School of Software, Northwestern Polytechnical University, Xi'an 710072, China. (e-mail: \{majunpeng, liyuke, jbwang\}@nwpu.edu.cn).}
\thanks{Yongxing Zhou is with Beijing University of Posts and Telecommunications, Beijing, China (email: yongxing.zhou@bupt.edu.cn).}
}

\markboth{Journal of \LaTeX\ Class Files,~Vol.~x, No.~x, April~2026}%
{Shell \MakeLowercase{\textit{et al.}}: A Sample Article Using IEEEtran.cls for IEEE Journals}

\maketitle

\begin{abstract}
Designing Doppler-resilient unimodular discrete phase-coded waveforms (DPWs) with low delay–Doppler sidelobes is critical for multiple-input multiple-output (MIMO) radar. Existing block coordinate descent (BCD) methods suffer from high computational cost for designing long sequences or large waveform sets. Meanwhile, learning-based alternatives such as the soft-quantization network (SQN) only address correlation optimization in the delay domain, without considering ambiguity function (AF) optimization in the joint delay–Doppler domain. To address these issues, this paper proposes a novel Doppler-resilient DPW design framework, termed SQNGD, for joint transmit–receive optimization that simultaneously optimizes the auto-AF, cross-AF (CAF), and signal-to-noise ratio loss (SNRL) under unimodular constraints. To solve the multi-objective optimization problem (MOOP), a joint transmit-receive design and an alternating optimization strategy are developed. The transmit waveforms are optimized via soft-quantization-based differentiable parameterization, while the receive filters are updated by gradient descent (GD) with an energy constraint and SNRL penalty. An FFT-accelerated evaluation of the AF and CAF is further incorporated, reducing the optimization time by $1.9\times$--$11\times$ compared with the state-of-the-art (SOTA) majorization–minimization-coordinate descent (MMCD) method. Numerical results show that SQNGD achieves a peak sidelobe level (PSL) of approximately $-43$~dB over the Doppler range $[-0.5,\,0.5]$ and $-31$~dB over $[-600,\,600]$, respectively, outperforming MMCD by $5.85$~dB and $3.45$~dB,  while 
maintaining the same SNRL of $0.5$~dB.

\end{abstract}


\section{Introduction}
In heavy clutter environments, such as autonomous driving scenarios, the active detection of small radar cross section (RCS) and low-velocity targets remains an intractable problem\cite{11184442}. For example, pedestrians exhibit weak radar cross sections and slow movement, posing serious challenges to automotive radar systems\cite{11408926}. The signal-dependent interference may come from roadside infrastructure,  guardrails, multipath reflections and illuminations from other radars\cite{11168247}. Due to the low RCS and small radial velocity of pedestrians, the signal-to-interference ratio is usually quite poor. In view of this, it is a key issue for active sensors to suppress the range–Doppler sidelobes of the probing waveform and prevent strong scatterers from masking weak targets. 
Besides, waveform design must deal with the notorious Doppler effect in high mobility environments. The ambiguity function (AF) provides a comprehensive characterization of waveform behavior in the delay–Doppler domain\cite{8903541,9724170,9018020} and is therefore regarded as a fundamental tool for characterizing waveform performance in this paper. Minimizing  range–Doppler sidelobes of AFs, particularly the peak sidelobe level (PSL), is essential for Doppler-resilient waveform design\cite{7469294,CHEN2023109075,8890867}.

Multiple-input multiple-output (MIMO) radar has emerged as a cornerstone technology for autonomous driving for its ability to deliver high-resolution range-Doppler-angle imaging with low hardware complexity\cite{11457727}.
Unimodular discrete phase-coded waveforms (DPWs) design has attracted significant attention in MIMO radars thanks to its realizability and robustness in hardware systems \cite{10128881,10348020,10620365,https://doi.org/10.1049/rsn2.12192,RAEI2023108914}. Compared with continuous phase-coded waveforms, DPWs can be readily implemented using low-resolution digital phase shifters. This property significantly reduces system complexity, power consumption, and calibration cost, making such waveforms highly practical for real-world applications\cite{7202844,8706639,7967829,9093027}.

Existing approaches to unimodular DPWs design can be broadly divided into two categories: transmitted-sequence design based on the matched-filter scheme, and joint transmit-receive design with transmit sequence and receive filter. For transmitter-only design methods, many works focus on optimizing the correlation properties of discrete-phase unimodular sequences, where low auto- and cross-correlation sidelobes are optimized to improve waveform orthogonality and weak-target detectability \cite{9745120,9440807}. For example, a block coordinate descent (BCD)-based method is developed in \cite{8979323} to design binary sequence sets for MIMO radar systems with good aperiodic and periodic auto- and cross-correlation properties. Beyond correlation optimization, other works also investigate AF shaping of waveforms to improve Doppler resilience by suppressing sidelobes over delay--Doppler regions of interest (ROI), such as majorization minimization and projected gradient descent algorithm (MM-PGD) based unimodular AF shaping method \cite{8844976}, fmincon-based method\cite{9404363}, and the CD framework for designing Doppler-sensitive binary and quaternary discrete-phase sequence sets for MIMO radars \cite{9409634}. Compared with transmitter-only design schemes, the joint transmit-receive design methods provide additional degrees of freedom by jointly designing the transmit waveforms and receive filters, therefore achieving lower range--Doppler sidelobes under a constrained signal-to-noise ratio (SNR) loss   \cite{8239836,7728070,10003252,rs15153877,9628070, 9765443, 9684877,9987663,10466432}. Representative methods include MM-Sequence Receiving Filter (MM-SR) and MM-Complementary Sequences Receiving Filters (MM-CSR)\cite{9765443}, weighted sidelobe level mismatched filter (WSL-MMF)\cite{9684877}, and several cyclic-algorithm-based methods\cite{9987663,10466432}. More recently, the majorization--minimization-coordinate descent (MMCD) method \cite{10679908} is proposed, where the transmit discrete-phase sequences are optimized via BCD and the receive filters are optimized by the MM algorithm. While these CD-based methods are widely adopted for DPWs design due to their guaranteed monotonic convergence and satisfactory sidelobe suppression, the computational complexity of sequential block-wise optimization increases rapidly as the length and number of waveforms increase, making them suffer from high computational cost for designing long sequences or large waveform sets.

Recently, learning-based optimization frameworks have emerged as a promising candidate for the DPW design, such as deep learning-based schemes  \cite{9257002,9618146}, long short-term memory (LSTM)-based binary code \cite{10453983},  model-based learned complex circle manifold network (LCCM-Net)\cite{10380119} and the soft-quantization network scheme (SQN)\cite{11079603}. Among these schemes,  the SQN introduces a differentiable approximation to discrete-phase constraints, enabling gradient-based optimization together with a soft-to-hard quantization strategy. Although SQN has demonstrated effectiveness in correlation sidelobe minimization, it is still limited to transmitter-only design based on the matched-filter scheme and does not address the complex issue of delay–Doppler sidelobe suppression. 


In this paper, a general joint design of Doppler-resilient DPWs and receiving filters framework, termed SQNGD, is developed for MIMO radars. In the proposed framework, the DPWs and receiving filters design problem is formulated as a multi-objective optimization problem (MOOP) to achieve the Pareto-optimal trade-off among AFs, cross-AFs (CAF), and the constrained SNRL with a unimodular constraint over the ROI. Note that the optimization of AFs and CAFs is much more challenging than conventional correlation optimization in SQN, since the sidelobe must be suppressed in both delay and Doppler domains, and the number of CAFs grows exponentially as the number of antennas or cross-channels increases. Besides, with discrete-phase constraints and SNRL control, the MOOP problem is inherently nonconvex and NP-hard, and the feasible search space grows exponentially with the sequence length and the phase alphabet size. Conventional gradient-based algorithms are not directly applicable to this problem, as the discrete feasible set introduces discontinuities that invalidate gradient computation.
To solve this MOOP problem, in the proposed framework the transmit waveforms are optimized through a soft-quantization-based differentiable parameterization inspired by SQN, while the receive filters are optimized via a gradient descent (GD)-based algorithm with an energy constraint. 
The contributions of this paper are summarized as follows:
\begin{itemize}
\item An optimized framework for designing Doppler-resilient DPWs and receiving filters, termed SQNGD, is proposed for MIMO radars. The designing problem is formulated as a MOOP to reach the Pareto-optimal trade-off among  three objectives, i.e., weighted auto-AF peak sidelobe level (WAPSL), the weighted CAF peak sidelobe level (WCPSL) and SNRL. Thereby, unimodular DPWs and receiving filters that achieve the desired AF and CAF shapes under constrained SNRL over different Doppler intervals of interest are obtained. 
Numerical results demonstrate that the proposed SQNGD achieves PSL of 
approximately $-43$~dB and $-31$~dB over the Doppler ranges 
$[-0.5,\,0.5]$ and $[-600,\,600]$, respectively, with improvements of 
$5.85$~dB and $3.45$~dB over the SOTA scheme MMCD, while 
maintaining the same SNRL of $0.5$~dB.
It should be mentioned that when the Doppler shift is set to zero, the proposed framework can also be used to jointly design waveforms and filters with desired auto- and cross-correlation properties. Numerical results show that SQNGD achieves the lowest correlation PSL with zero SNRL, providing almost $3$~dB and $1$~dB gains compared with the SQN and the MMCD, respectively.


\item A joint transmit-receive design  and an alternating optimization strategy are designed to solve the aforementioned MOOP. On the transmit side, inspired by SQN, designing DPWs is optimized through the soft-quantization-based differentiable parameterization, while on the receive side, the receiving filters are optimized by GD with the energy constraint. To coordinate the two parts, an alternating optimization strategy and a stopping criterion are developed for the joint design of transmit waveforms and receive filters. The proposed framework provides a more flexible approach for designing DPWs and receiving filters. On the one hand, the relative priorities among the three sub-objectives, i.e., WAPSL, WCPSL, and SNRL, can be flexibly adjusted to suit specific application requirements. On the other hand, the waveform parameters, such as the sequence length of DPWs, the phase alphabet size, and the range of the Doppler interval, can be customized in the proposed framework.

\item To reduce computational complexity and enable efficient solution of the MOOP, two acceleration methods are incorporated into the proposed framework. First, the SQNGD employs FFT-based batch gradient updates, replacing the sequential coordinate-wise discrete search used in conventional CD algorithms. This design is inherently more amenable to parallelization and acceleration, and the runtime advantage is expected to become increasingly pronounced as the problem scale grows. Second, an FFT-accelerated mode is leveraged to replace the computationally intensive correlation operations involved in calculating the AFs of the loss function. Experimental results demonstrate that the proposed approaches reduce the optimization time by at least $1.9$× compared to the MMCD algorithm, with the speedup reaching up to $11$×  as the sequence length and the phase alphabet size increase.



\end{itemize}

\textit{Notation:} Throughout this paper, boldface lowercase letters denote column vectors and boldface uppercase letters denote matrices. $(\cdot)^T$, $(\cdot)^*$, and $(\cdot)^H$ denote the transpose, complex conjugate, and conjugate transpose, respectively. $\|\cdot\|_2$ denotes the Euclidean norm, and $|\cdot|$ denotes the absolute value. $\mathbb{C}$ and $\mathbb{R}$ denote the complex and real number fields. $\mathrm{vec}(\cdot)$ denotes the vectorization operator, and $\mathrm{mat}(\cdot)$ denotes its inverse reshape operation. $\odot$ denotes the Hadamard product. $\mathrm{FFT}(\cdot)$ and $\mathrm{IFFT}(\cdot)$ denote the discrete Fourier transform and its inverse. $j=\sqrt{-1}$ denotes the imaginary unit.

\section{PROBLEM STATEMENT}

\subsection{Unimodular DPWs Design}

Consider a MIMO radar system with $M$ transmit antennas.
Each antenna transmits a unimodular discrete-phase sequence of length $N$, denoted by
\begin{equation}
\mathbf{x}_m = [x_m(1), \ldots, x_m(n), \ldots, x_m(N)]^T \in \mathbb{C}^N,
\end{equation}
where $m = 1,\ldots,M$, $n = 1,\ldots,N$, $\mathbb{C}$ is a complex space, and $(\cdot)^T$ denotes the transpose operation. Each sequence element is constructed as
\begin{equation}
x_m(n) = e^{j y_m(n)},
\end{equation}
where $y_m(n)$ denotes the phase of the transmitted signal, and $x_m(n)$ satisfies the unimodular constraint.

Stacking all transmit sequences yields the waveform matrix
\begin{equation}
\mathbf{X} = [\mathbf{x}_1, \ldots, \mathbf{x}_m, \ldots, \mathbf{x}_M] \in \mathbb{C}^{N \times M},
\end{equation}
and the corresponding phase matrix
\begin{equation}
\mathbf{Y} = [\mathbf{y}_1, \ldots, \mathbf{y}_m, \ldots, \mathbf{y}_M] \in \mathbb{R}^{N \times M},
\end{equation}
where $\mathbb{R}$ is a real space.
Assuming that discrete phase coding with phase alphabet size $B$ is employed, the phase variables satisfy
\begin{equation}
y_m(n) \in \left\{ \frac{2\pi b}{B} \;\big|\; b = 0,1,\ldots,B-1 \right\}.
\end{equation}
Let $\mathbf{H} = [\mathbf{h}_1 ,\ \mathbf{h}_2, \ \cdots ,\ \mathbf{h}_M]\in \mathbb{C}^{N \times M}$ denote the corresponding receive filters, which satisfy the energy
constraint $\|\mathbf{h}_m\|_2^2=N$, where $\|\cdot\|_2$ denotes the Euclidean norm.
To characterize waveform orthogonality in scenarios involving target motion, the AF is adopted as the performance metric. Unlike conventional correlation functions defined only in the delay domain, the AF characterizes waveform behavior jointly in the delay–Doppler domain, providing a more comprehensive description of range--Doppler resolution and sidelobe interference.

The discrete AF between the transmit waveform $\mathbf{x}_m$ and the receive filter $\mathbf{h}_l$ at delay $\tau$ and normalized Doppler frequency $f$ is defined as
\begin{equation}
A_{ml}(\tau,f)=
\begin{cases}
\displaystyle \sum_{n=1}^{N-\tau} x_m(n)\, h_l^*(n+\tau)\,
e^{j 2\pi n \frac{f}{N}}, & 0 \le \tau, \\[8pt]
\displaystyle \sum_{n=-\tau+1}^{N} x_m(n)\, h_l^*(n+\tau)\,
e^{j 2\pi n \frac{f}{N}}, &  \tau < 0,
\end{cases}
\label{eq:AF}
\end{equation}
where $m,l=1,\ldots,M$, $-N+1 \le \tau \le N-1$, $f \in [-N/2,\,N/2]$, and $(\cdot)^*$ denotes complex conjugation. The normalized Doppler variable $f$ is defined as $f = f_d T,$
where $f_d$ denotes the Doppler frequency shift and $T$ is the waveform duration. When $m=l$, (6) corresponds to the aperiodic AF, whereas for $m \neq l$ it represents the aperiodic CAF.

This paper focuses on minimizing the worst weighted PSL (WPSL) over the delay--Doppler ROI, which is defined as
\begin{equation}
\mathrm{WPSL} =
\max \left\{
\text{WAPSL},\;
\text{WCPSL}
\right\}.
\end{equation}
Accordingly, the WPSL can be decomposed into the WAPSL and the WCPSL, given by
\begin{equation}
\text{WAPSL} = \max_{m_1= m_2, \tau \neq 0} \left\{ w(\tau,f) \left| A_{x_{m_1} x_{m_2}}(\tau, f) \right| \right\},
\end{equation}
and
\begin{equation}
\text{WCPSL} = \max_{m_1 \neq m_2} \left\{ w(\tau,f) \left| A_{x_{m_1} x_{m_2}}(\tau, f) \right| \right\},
\end{equation}
respectively. $w(\tau, f)$ is a weighting coefficient that can be tailored to emphasize specific delay–Doppler regions.
Waveforms with low APSL can effectively suppress range-Doppler sidelobes in the AF, thereby improving weak target detection performance in the presence of Doppler shifts. Meanwhile, low CPSL reduces mutual interference among different transmit channels, which enhances waveform orthogonality and improves target parameter estimation accuracy in MIMO radar systems.

The SNRL due to mismatched filtering is defined as
\begin{equation}
\mathrm{SNRL}
=10\log_{10}\left(
\frac{\|x_m\|_2^2\|h_m\|_2^2}{|x_m^H h_m|^2}
\right),
\label{eq:SNRL}
\end{equation}
and $(\cdot)^H$ denotes the conjugate transpose.
A lower SNRL indicates a smaller performance degradation relative to the matched filter benchmark.
\subsection{Problem Formulation}

In this paper, the waveform design objective is to suppress the worst WPSL of the AF while controlling SNRL. 
Therefore, the joint optimization problem of DPWs and receive filters for AF shaping under the SNRL constraint can be formulated as:
\begin{equation}\label{eq:opt_problem}
\begin{aligned}
\min_{\mathbf{X},\mathbf{H}} \quad &
\tilde{\Gamma} =
\varepsilon\, \mathrm{WPSL} +
(1-\varepsilon)\, \tilde{G} \\
\text{s.t.} \quad &
x_m(n) = e^{j y_m(n)}, \quad
y_m(n) \in \left\{ \tfrac{2\pi b}{B} \right\}_{b=0}^{B-1}, \\
&\|\mathbf{h}_m\|_2^2=N,\quad m=1,\ldots,M,\; n=1,\ldots,N ,
\end{aligned}
\end{equation}
where $\varepsilon \in [0,1]$ is a weighting coefficient that balances sidelobe suppression and SNR preservation.

The auxiliary term $\tilde{G}$ imposes a mainlobe constraint and is defined as
\begin{equation}\label{eq:g_tilde}
\tilde{G}
=
\sum_{m=1}^{M}
\left|
h_m^H x_m - \tilde{a}_{\max}
\right|^2,
\end{equation}
where $\tilde{a}_{\max}$ denotes the target zero-delay and zero-Doppler mainlobe amplitude. To achieve a desired SNRL $\tilde{\mu}$ (in dB), define the target normalized mainlobe amplitude as
\begin{equation}\label{eq:ptar_main}
p_{\mathrm{tar}} = 10^{-\tilde{\mu}/20}.
\end{equation}
Accordingly, the target mainlobe amplitude is given by
\begin{equation}\label{eq:amax_main}
\tilde{a}_{\max} = N\,p_{\mathrm{tar}} = N\,10^{-\tilde{\mu}/20}.
\end{equation}

The MOOP problem  (\ref{eq:opt_problem}) is non-convex due to the unimodular and the discrete phase constraints. Existing BCD approaches typically address this problem by sequentially updating block phase entries, which leads to high computational complexity. This limitation motivates the development of the SQNGD optimization framework.

\section{THE PROPOSED OPTIMIZATION FRAMEWORK}

\begin{figure*}[!t]
  \centering
  \includegraphics[width=\textwidth]{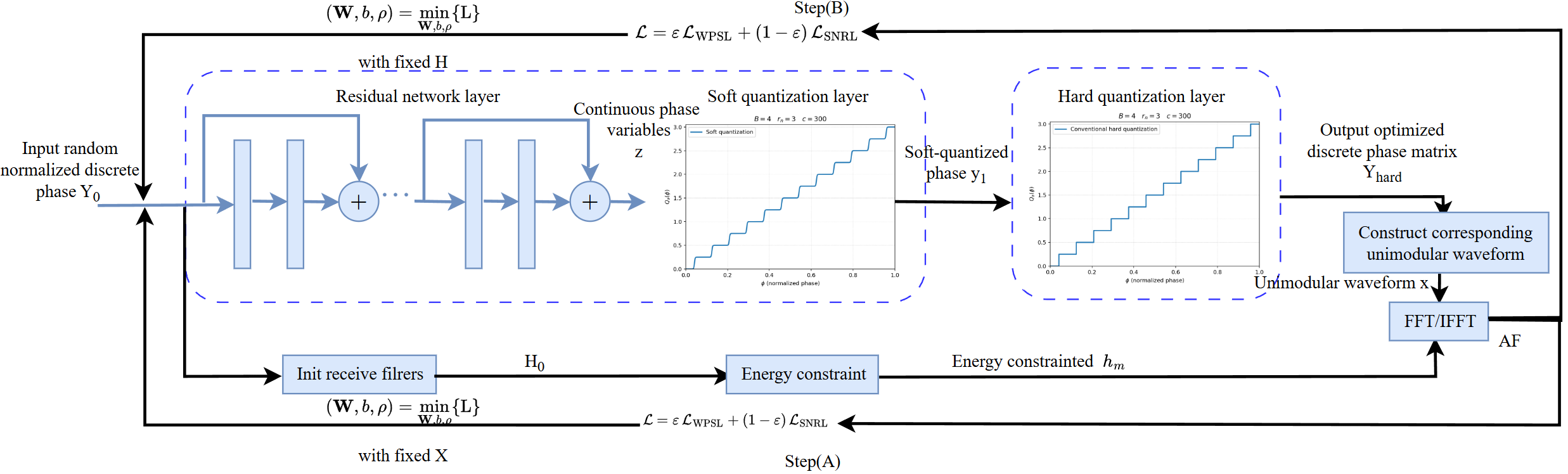}
  \caption{The SQNGD framework.}
  \label{fig:SQNGDframework}
\end{figure*}

In this section, the SQNGD framework is proposed for joint transmit--receive design under discrete-phase constraints, as illustrated in Fig.~\ref{fig:SQNGDframework}. The proposed framework is designed to minimize the WPSL in the delay--Doppler domain by jointly optimizing the discrete-phase transmit waveforms and the energy-constrained receive filters. In this framework, the discrete-phase constraint at the transmitter is handled through a soft-quantization mechanism, enabling gradient-based optimization. Meanwhile, the receive filters are modeled as trainable complex-valued variables and optimized under an energy constraint.

To achieve improved sidelobe suppression and enhanced optimization stability, an alternating optimization strategy is adopted to solve the joint design problem. Specifically, with the hard-quantized transmit waveforms fixed, the receive filters are updated to suppress delay--Doppler sidelobes while satisfying the energy constraint. Subsequently, with the receive filters fixed, the transmit-side parameters are updated using soft-quantized waveforms through differentiable AF computation. After convergence, a hard quantization stage is applied to generate valid discrete-phase transmit waveforms.

\subsection{SQNGD Design}

Following the waveform model in Section~II, the phase matrix is denoted by $\mathbf{Y}\in\mathbb{R}^{N\times M}$. For ease of optimization, $\mathbf{Y}$ is first normalized as
\begin{equation}
\mathbf{Y}_0=\frac{\mathbf{Y}}{2\pi},
\end{equation}
where $\mathbf{Y}_0\in\mathbb{R}^{N\times M}$ is the normalized phase matrix, and each elements is selected from $[0,1]$. $\mathbf{Y}_0$ is then reshaped into a vector $\mathbf{y}_0$ as
\begin{equation}
\mathbf{y}_0=\mathrm{vec}(\mathbf{Y}_0)\in\mathbb{R}^{NM},
\end{equation}
and serves as the input to SQNGD.


A deep residual network (ResNet) is employed to generate continuous phase variables $z$ as
\begin{equation}
\mathbf{z} = f_{\boldsymbol{\theta}}(\mathbf{y}_0),
\end{equation}
where $f_{\boldsymbol{\theta}}(\cdot)$ denotes the ResNet parameterized by network parameters $\boldsymbol{\theta}$. The output $\mathbf{z}$ is continuous and does not satisfy the discrete-phase constraint; therefore, it must be quantized.

The ideal hard quantization operator maps a continuous variable to the nearest discrete level:
\begin{equation}
Q_R^{'}(x) = \arg\min_{q \in \mathcal{Q}} |x - q|,
\quad
\mathcal{Q} = \left\{ \frac{b}{B} \right\}_{b=0}^{B-1},
\end{equation}
where $|\cdot|$ denotes the absolute value. However, $Q_R^{'}(\cdot)$ is non-differentiable, which prevents the use of gradient-based optimization.

To enable backpropagation, SQNGD introduces a differentiable approximation of $Q_R'(\cdot)$ using the tanh function. The tanh function exhibits a smooth yet step-like transition and satisfies
\begin{equation}
\tanh(x) \in(-1,1)
\end{equation}
and
\begin{equation}
\frac{d}{dx}\tanh(x) = 1 - \tanh^2(x) \in (0,1],
\end{equation}
which provides bounded and stable gradients.

\begin{figure}[!t]
  \centering
  \includegraphics[width=\columnwidth]{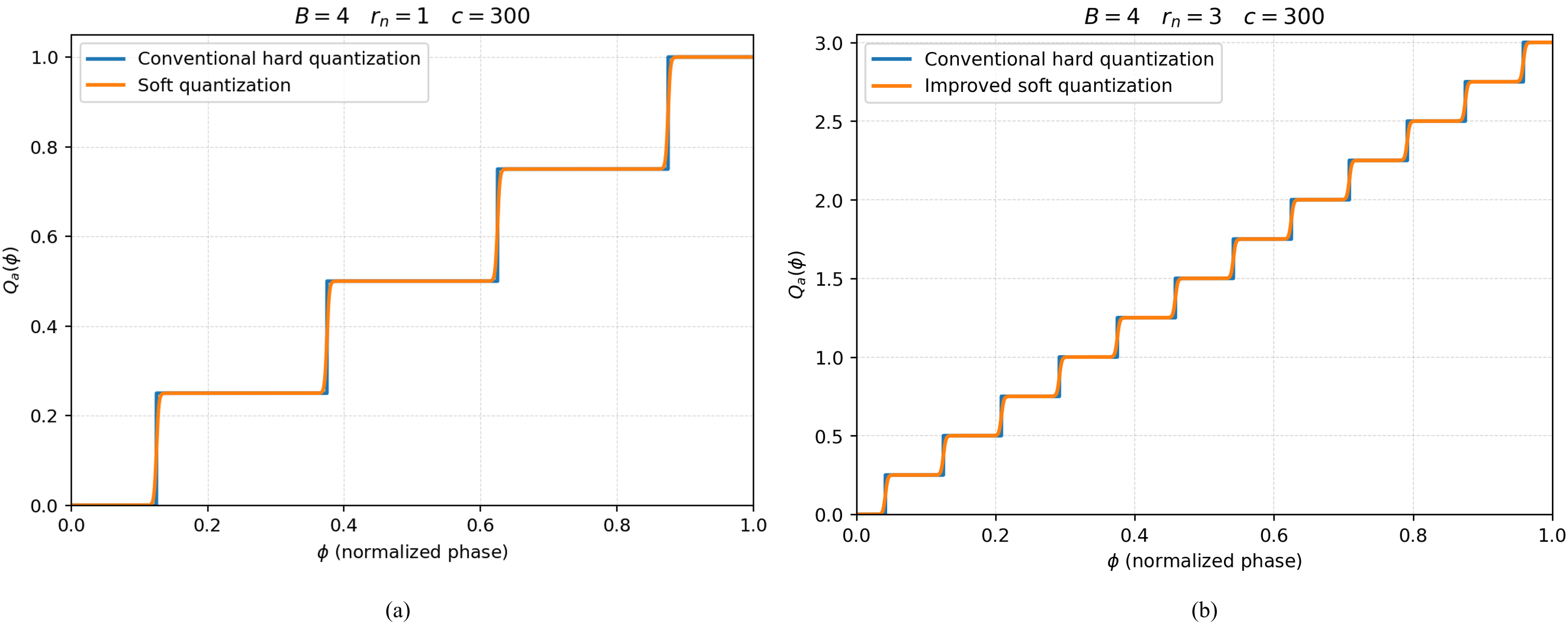}
  \caption{The soft quantization function used for training.
(a) The original soft quantization function.
(b) The improved soft quantization function.}
  \label{fig:SQN}
\end{figure}
The soft quantization operator is constructed as
\begin{equation}
Q_A(x)
=
\sum_{i=0}^{B-1}
a_i
\big[
\tanh\big(c(x - \rho_i)\big) + 1
\big],
\end{equation}
where $c>0$ is a hyperparameter controlling the sharpness of the soft transition, $\rho_i$ are quantization thresholds, and $a_i$ are scaling coefficients associated with the $i$-th transition, as shown in Fig.~\ref{fig:SQN} (a). Compared with fixed-threshold quantization, $\{\rho_i\}$ can be treated as trainable parameters and optimized jointly with the network to reduce quantization mismatch.

The soft-quantized phase vector is given by
\begin{equation}
\mathbf{y}_1 = Q_A(\mathbf{z}),
\qquad
\mathbf{Y}_1 = 2\pi \cdot \mathrm{mat}(\mathbf{y}_1),
\end{equation}
where $\mathrm{mat}(\cdot)$ denotes the inverse operation of $\mathrm{vec}(\cdot)$, i.e., it reshapes the vector $\mathbf{y}_1$ into the matrix form $\mathbf{Y}_1 \in \mathbb{R}^{N\times M}$. The corresponding unimodular waveform used for AF evaluation is then given by
\begin{equation}
x_m(n) = e^{j2\pi y_{1,m}(n)}.
\end{equation}

When the parameter $c$ increases, the soft quantization layer approaches the hard quantization layer. However, an excessively large $c$ may degrade the training performance in practice. Inspired by BCD, reducing the spacing between the quantization thresholds $\rho_i$ allows the network to explore more combinations of phase sequences. To make the soft quantization layer closer to the hard quantization layer while maintaining good optimization performance, a copying mechanism (see Fig.~\ref{fig:SQN} (b)) is adopted.
Specifically, the number of copies is set to $r_n$, and $r_n$ soft quantization layers defined from 0 to 1, denoted as $[0,\,1]$, are superimposed and compressed back to $[0,\,1]$, which effectively increases the number of quantization regions from $B$ to $B\times r_{n}$. The resulting soft quantization model is given by
\begin{equation}
Q_A(x)=\sum_{i=0}^{B\times r_{n}-1} a_i\big[\tanh\big(c(x-\rho_i)\big)+1\big].
\label{eq:QA_copy}
\end{equation}

After hard quantization, the obtained normalized values are mapped to phases. For example, in the quad-phase case with $B=4$, $r_n=3$ and $c=300$, the hard-quantized results correspond to
\begin{equation}
2\pi \times\left\{ \frac{k}{4} \,\middle|\, k=0,1,\ldots,12 \right\}.
\end{equation}
Since the phase has a period of $2\pi$, the final waveform remains a four-phase encoding, and thus the discrete-phase constraint is not violated. Nevertheless, the copying mechanism enlarges the search space during training and improves the optimization performance compared with the case without the copying mechanism.


Based on the multichannel receive filters $\mathbf{h}_m$, the WPSL of AF is evaluated jointly with the
discrete transmit waveforms. To enable gradient-based optimization while preserving the complex structure, the receive filters
are modeled as trainable complex-valued variables with the energy constraint.

For an $M$-channel system with waveform length $N$, define
the receive-filter matrix
\begin{equation}
\mathbf{H} = [\mathbf{h}_1 ,\ \mathbf{h}_2, \ \cdots ,\ \mathbf{h}_M]\in \mathbb{C}^{N \times M},
\end{equation}
where $\mathbf{h}_m \in \mathbb{C}^{N}$ denotes the filter associated with the $m$-th channel. In practice, the receive-filter matrix $\mathbf{H}$ is parameterized by two trainable real-valued matrices,
\begin{equation}
\mathbf{H}_{\text{Re}} \in \mathbb{R}^{N\times M}, \qquad
\mathbf{H}_{\text{Im}} \in \mathbb{R}^{N\times M},
\end{equation}
where $\mathbf{H}_{\text{Re}}$ and $\mathbf{H}_{\text{Im}}$ denote the real and imaginary parts of $\mathbf{H}$, respectively. The complex receive-filter matrix is then reconstructed as
\begin{equation}\label{eq:H_reim}
\mathbf{H} = \mathbf{H}_{\text{Re}} + j\,\mathbf{H}_{\text{Im}}.
\end{equation}

The receive filters are initialized using the matched filter from the initial unimodular transmit
sequence. Let $\mathbf{Y}_{0}$ denote the initial normalized phase matrix. 
The initial transmit waveform is then defined as
\begin{equation}
\mathbf{X}_{0}=\exp\!\big(j2\pi \mathbf{Y}_{0}\big)\in\mathbb{C}^{N\times M},
\end{equation}
applied element-wise. The initial receive filters are set to the conjugate of the transmit waveform,
\begin{equation}
\mathbf{H}_0 = \mathbf{X}_0^{*}.
\end{equation}
where $(\cdot)^{*}$ denotes the complex conjugate operation. The trainable parameters are then initialized as
$\mathbf{H}_{\text{Re}}=\mathrm{Re}(\mathbf{H}_0)$ and $\mathbf{H}_{\text{Im}}=\mathrm{Im}(\mathbf{H}_0)$.

To control receive-filter gain and ensure consistent scaling across channels, the energy constraint
\begin{equation}
\|\mathbf{h}_m\|_2^2 = N,\qquad m = 1,\ldots,M,
\label{eq:H_energy}
\end{equation}
is enforced throughout training. Let $\mathcal{H} = \left\{ \mathbf{h} \in \mathbb{C}^{N} \mid \|\mathbf{h}\|_{2}^{2} = N \right\}$ denote the energy-constrained feasible set. Define the projection onto $\mathcal{H}$ as
\begin{equation}
\Pi_{\mathcal{H}}(\mathbf{h}_m)
=
\sqrt{\frac{N}{\|\mathbf{h}_m\|_{2}^{2}}}\;\mathbf{h}_m,
\qquad m = 1, \ldots, M.
\label{eq:H_proj}
\end{equation}
After each gradient update, each receive filter is replaced by its projected value,
$\mathbf{h}_m = \Pi_{\mathcal H}(\mathbf{h}_m)$.
This projection normalizes the energy of each receive filter to $N$.

With the transmit waveform fixed, the receive
filters are optimized by minimizing the WPSL objective with respect to the trainable variables in
\eqref{eq:H_reim}. For each outer iteration, $N_h$ inner steps of Adam are performed on
$(\mathbf{H}_{\text{Re}}, \mathbf{H}_{\text{Im}})$ using the gradient of the joint loss, and the projection \eqref{eq:H_proj} is
applied after every Adam step to maintain feasibility under \eqref{eq:H_energy}. Consequently, $\mathbf{H}$ remains energy-constrained throughout the optimization process while adapting to suppress delay--Doppler sidelobes under the prescribed objective.

\subsection{FFT Accelerated AF Computation and Loss Function with SNRL Penalty}\label{subsec:joint_af}

Based on the jointly designed transmit waveforms $\mathbf{x}_m$ and receive filters $\mathbf{h}_\ell$, the AF is evaluated according to the formulation in Section~II. To obtain an efficient and differentiable implementation for joint optimization, the AF is computed via FFT-accelerated aperiodic correlation for each Doppler bin. This formulation supports backpropagation with respect to both the transmit-phase variables and the receive filters $\mathbf{h}_\ell$.

For a fixed Doppler bin $f$, define the Doppler-modulated transmit waveform
\begin{equation}
\hat{x}_{m,f}(n)=x_m(n)\,e^{j2\pi n f/N}, \quad n=1,\ldots,N,
\end{equation}
where $\hat{x}_{m,f}(n)$ denotes the $n$-th sample of the $m$-th transmit waveform after Doppler modulation at Doppler bin $f$. Then the AF between the $m$-th transmit channel and the $\ell$-th receive channel can be written as the aperiodic
correlation
\begin{equation}
\label{eq:af_joint_time}
A_{m\ell}(\tau,f)=
\begin{cases}
\displaystyle \sum_{n=1}^{N-\tau}\hat{x}_{m,f}(n)\,h_\ell^*(n+\tau), & 0 \le \tau, \\[6pt]
\displaystyle \sum_{n=1-\tau}^{N}\hat{x}_{m,f}(n)\,h_\ell^*(n+\tau), & \tau < 0,
\end{cases}
\end{equation}
where $A_{m\ell}(\tau,f)$ denotes the AF between the $m$-th transmit channel and the $\ell$-th receive channel at delay $\tau$ and Doppler bin $f$, and $-N+1 \le \tau \le N-1$.

The correlation in \eqref{eq:af_joint_time} can be computed efficiently using FFT with zero padding. Specifically, let
\begin{equation}
\hat{\mathbf{x}}^{\mathrm{pad}}_{m,f}=
\big[\hat{\mathbf{x}}_{m,f}^T,\ \mathbf{0}_{N-1}^T\big]^T,\qquad
\mathbf{h}^{\mathrm{pad}}_{\ell}=
\big[\mathbf{h}_{\ell}^T,\ \mathbf{0}_{N-1}^T\big]^T,
\end{equation}
where $\hat{\mathbf{x}}^{\mathrm{pad}}_{m,f}$ and $\mathbf{h}^{\mathrm{pad}}_{\ell}$ are the corresponding zero-padded vectors, and $\mathbf{0}_{N-1}$ denotes an all-zero vector of length $N-1$. The correlation sequence is obtained as
\begin{equation}\label{eq:fft_corr_joint}
\mathbf{r}_{m\ell,f}
=
\mathrm{IFFT}\!\left(
\mathrm{FFT}\!\left(\hat{\mathbf{x}}^{\mathrm{pad}}_{m,f}\right)
\odot
\mathrm{conj}\!\left(\mathrm{FFT}\!\left(\mathbf{h}^{\mathrm{pad}}_{\ell}\right)\right)
\right),
\end{equation}
where $\odot$ denotes the Hadamard product and $\mathrm{FFT}(\cdot)$ and $\mathrm{IFFT}(\cdot)$ denote the discrete Fourier
transform and its inverse.

Since the above procedure consists of Doppler modulation, FFT and IFFT operations and element-wise complex multiplication, the AF computation can be embedded as a differentiable module and support backpropagation with
respect to the transmit waveforms $\mathbf{x}_m$ and the receive filters $\mathbf{h}_\ell$, enabling joint transmit--receive optimization under the WPSL objective.


In this paper, the AF-shaping objective is to suppress the worst WPSL of the AF while
controlling the SNRL. Accordingly, a WPSL loss is adopted to penalize the peak sidelobes, and an auxiliary penalty term is incorporated to regulate the zero-delay mainlobe peak.

Let $\mathrm{WAPSL}$ and $\mathrm{WCPSL}$ denote the auto- and cross-AF peak sidelobe level metrics in specific
delay–Doppler ROI, respectively. The sidelobe loss is constructed as
\begin{equation}\label{eq:L_psl}
\mathcal{L}_{\mathrm{WPSL}}
=
\max\!\left(
\mathrm{WAPSL},\,
\mathrm{WCPSL}
\right).
\end{equation}
By minimizing $\mathcal{L}_{\mathrm{WPSL}}$, the maximum sidelobe of both the AFs and CAFs is suppressed
over the ROI, which improves Doppler-resilient sidelobe performance under the discrete-phase
constraint.

To control the SNRL, an auxiliary mainlobe penalty is imposed on the zero-delay and zero-Doppler mainlobe. Specifically, define
\begin{equation}\label{eq:mainlobe_ratio}
p_m
=
\frac{|h_m^H x_m|}{N}
=
\frac{|A_{mm}(0,0)|}{N},
\qquad m=1,\ldots,M,
\end{equation}
where $p_m$ denotes the normalized mainlobe amplitude of the $m$-th transmit waveform–receive filter pair. Let the target normalized mainlobe amplitude be denoted by $p_{\mathrm{tar}}$, which is determined by the desired SNRL $\tilde{\mu}$ (in dB) as
\begin{equation}\label{eq:ptar}
p_{\mathrm{tar}} = 10^{-\tilde{\mu}/20}.
\end{equation}
Accordingly, the target mainlobe amplitude can be written as
\begin{equation}\label{eq:amax_tar}
\tilde{a}_{\max}=N\,p_{\mathrm{tar}},
\end{equation}
which is consistent with the mainlobe constraint in \eqref{eq:g_tilde}. The corresponding SNRL penalty term is then defined as
\begin{equation}\label{eq:L_snrl}
\mathcal{L}_{\mathrm{SNRL}}
=
\frac{1}{M}\sum_{m=1}^{M}
\left(
p_m - p_{\mathrm{tar}}
\right)^2 .
\end{equation}

Finally, the overall loss function with SNRL penalty is formulated as
\begin{equation}\label{eq:L_total}
\mathcal{L}
=
\varepsilon\,\mathcal{L}_{\mathrm{WPSL}}
+
(1-\varepsilon)\,\mathcal{L}_{\mathrm{SNRL}},
\end{equation}
where $\varepsilon\in[0,1]$ is a weighting coefficient that balances sidelobe suppression and SNR preservation. This loss encourages low WPSL over the delay--Doppler ROI while maintaining the desired mainlobe level at $\tau=0$, thereby controlling the SNRL in
the joint transmit--receive design.

\subsection{Alternating Optimization Strategy and Stopping Criterion}
Problem~\eqref{eq:opt_problem} is nonconvex due to the bilinear coupling between the transmit waveforms and receive
filters in the AF, as well as the discrete-phase constraint. An alternating optimization strategy is
therefore adopted by partitioning the variables into two blocks, i.e., the receive filters $\mathbf{H}$ and the
transmit-side SQN parameters that generate $\mathbf{X}$.

In the $t$-th outer iteration, a hard-quantized unimodular waveform $\mathbf{X}^{(t)}_{\mathrm{hard}}$ is first
generated from the current phase parameters and then treated as a constant (i.e., no gradient is propagated through hard-quantization path). With $\mathbf{X}^{(t)}_{\mathrm{hard}}$ fixed, the receive filters are updated by
performing $N_h$ steps of Adam on $\mathbf{H}$ to minimize the joint objective involving WPSL and SNRL, followed by a row-wise
projection $\mathbf{h}_m = \Pi_{\mathcal H}(\mathbf{h}_m)$ after each step to enforce the energy constraint.

Next, the updated receive filters $\mathbf{H}^{(t+1)}$ are held fixed and the transmit waveform parameters are updated
by performing $N_x$ steps of Adam on the SQN parameters using the \emph{soft}-quantized waveform
$\mathbf{X}^{(t)}_{\mathrm{soft}}$, which preserves differentiability and enables backpropagation through the AF
computation. After each transmit-side update, the quantization thresholds are projected to maintain a valid ordered
set. These two steps are repeated for $t=1,\ldots,T$, yielding an alternating optimization procedure that is scalable
to large $N$ and dense delay--Doppler sampling.


After training, the hard quantization layer is applied to map the continuous network output $\mathbf{y}_1$ to a discrete representation. Let $\{\rho_i\}_{i=0}^{B\times r_n-1}$ denote the ordered quantization thresholds satisfying $\rho_0<\rho_1<\cdots<\rho_{B\times r_n-1}$. The hard quantization operator is defined as
\begin{equation}
Q_R(x)=
\begin{cases}
0, & x<\rho_0,\\[2mm]
Q_A\!\left(\dfrac{\rho_i+\rho_{i+1}}{2}\right), & \rho_i \le x < \rho_{i+1},\ \\[2mm]
\sum_{i=0}^{B\times r_n-1} 2a_i, & \rho_{B\times r_n-1}\le x.
\end{cases}
\label{eq:QR_piecewise}
\end{equation}
The corresponding DPWs are then generated by phase mapping, and the hard WPSL is evaluated using the same formulation as (7).

Training terminates when further reduction of $\mathcal{L}$ does not lead to improvement in the WPSL or the maximum number of outer iterations $T$ is reached.

\begin{algorithm}[!t]
\caption{SQNGD for Joint Transmit--Receive Design}
\label{alg:SQNGD_joint}
\begin{algorithmic}[1]
\State \textbf{Input:} $M,N,B$; Doppler grid $\mathcal{F}$; ROI weights $w(\tau,f)$; 
weighting coefficient $\varepsilon\in[0,1]$; target SNRL $\tilde{\mu}$ (dB) with $p_{\mathrm{tar}}=10^{-\tilde{\mu}/20}$; 
inner steps $N_h,N_x$; learning rates $\eta_h,\eta_x$; outer iterations $T$.

\State \textbf{Initialization:}
\Statex \quad Random
normalized discrete
phase $\mathbf{Y}_0$ and set $\mathbf{y}_0=\mathrm{vec}(\mathbf{Y}_0)$.
\Statex \quad $\mathbf{X}_0= \exp\!\big(j2\pi\mathbf{Y}_0\big)$, \ \ $\mathbf{H}= \mathrm{conj}(\mathbf{X}_0)$.
\Statex \quad Project rows: $\mathbf{h}_m = \Pi_{\mathcal H}(\mathbf{h}_m),\ \forall m$.
\Statex \quad Initialize SQN parameters $(\boldsymbol{\theta},\boldsymbol{\rho})$.

\For{$t=1$ to $T$}
    \State \textbf{Waveform generation:}
    \Statex \quad $\mathbf{z}= f_{\boldsymbol{\theta}}(\mathbf{y}_0)$.
    \Statex \quad $\mathbf{y}_{\mathrm{soft}}= Q_A(\mathbf{z};\boldsymbol{\rho})$, \ \ $\mathbf{y}_{\mathrm{hard}}= Q_R(\mathbf{z};\boldsymbol{\rho})$.
    \Statex \quad $\mathbf{X}_{\mathrm{soft}}= \exp\!\big(j2\pi\,\mathrm{mat}(\mathbf{y}_{\mathrm{soft}})\big)$.
    \Statex \quad $\mathbf{X}_{\mathrm{hard}}= \exp\!\big(j2\pi\,\mathrm{mat}(\mathbf{y}_{\mathrm{hard}})\big)$.

    \State \textbf{Step (A): Update $\mathbf{H}$ with fixed $\mathbf{X}_{\mathrm{hard}}$:}
    \For{$i=1$ to $N_h$}
        \State Compute AF $\mathbf{A}(\mathbf{X}_{\mathrm{hard}},\mathbf{H})$.
        \State $\mathcal{L}_{\mathrm{WPSL}}= \max(\mathrm{WAPSL},\mathrm{WCPSL})$.
        \State 
        $\mathcal{L}_{\mathrm{SNRL}}
        =
        \frac{1}{M}\sum_{m=1}^{M}
        \left(
        p_m - p_{\mathrm{tar}}
        \right)^2 .$
        \State $\mathcal{L}= \varepsilon\,\mathcal{L}_{\mathrm{WPSL}}+(1-\varepsilon)\,\mathcal{L}_{\mathrm{SNRL}}$.
        \State Update $\mathbf{H}$ by Adam using $\nabla_{\mathbf{H}}\mathcal{L}$.
        \State Energy constraint: $\mathbf{h}_m = \mathbf{h}_m\sqrt{\frac{N}{\|\mathbf{h}_m\|_2^2}}$, $\forall m$.
    \EndFor

    \State \textbf{Step (B): Update SQN parameters with fixed $\mathbf{H}$:}
    \For{$j=1$ to $N_x$}
        \State Compute AF $\mathbf{A}(\mathbf{X}_{\mathrm{soft}},\mathbf{H})$ .
        \State Compute $\mathcal{L}_{\mathrm{WPSL}}$, $\mathcal{L}_{\mathrm{SNRL}}$, and $\mathcal{L}$ as above.
        \State Update $(\boldsymbol{\theta},\boldsymbol{\rho})$ by Adam using $\nabla_{(\boldsymbol{\theta},\boldsymbol{\rho})}\mathcal{L}$.
    \EndFor

\State \textbf{Stopping criterion:} Training terminates when further reduction of $\mathcal{L}$ does not lead to improvement in the WPSL, or when the maximum number of outer iterations $T$ is reached.
\EndFor

\State \textbf{Output:} $(\mathbf{X}_{\mathrm{hard}},\mathbf{H})$.
\end{algorithmic}
\end{algorithm}

\textbf{Remarks:}
The computational cost of SQNGD mainly comes from the FFT-accelerated AF evaluation and the forward and backward propagation of the SQN. Since the AF is evaluated over all $F$ Doppler bins and all $M^2$ transmit--receive channel pairs, the complexity for each AF evaluation is
\begin{equation}
\mathcal{C}_{\mathrm{AF}}
=
\mathcal{O}\!\left(M^2FN\log N\right).
\end{equation}
where $\mathcal{C}_{\mathrm{AF}}$ denotes the computational complexity of each AF evaluation, and $\mathcal{O}(\cdot)$ denotes the big-O order notation.
Accordingly, the leading complexity per outer iteration of SQNGD is
\begin{equation}
\mathcal{C}_{\mathrm{SQNGD}}
=
\mathcal{O}\!\left((N_h+N_x)M^2FN\log N + N_x\mathcal{C}_{\mathrm{SQN}}\right).
\end{equation}

For MMCD, updating a single waveform entry requires $2N-1$ computations of a $B$-point DFT, each with complexity $\mathcal{O}(B\log_2 B)$. Thus, the complexity of each waveform-entry update is
\begin{equation}
\mathcal{C}_{\mathrm{entry}}^{\mathrm{MMCD}}
=
\mathcal{O}(2NB\log_2 B),
\end{equation}
and the dominant transmit-side complexity per outer iteration is
\begin{equation}
\mathcal{C}_{\mathrm{MMCD}}
=
\mathcal{O}(2MN^2B\log_2 B).
\label{eq:complexity_mmcd}
\end{equation}

Therefore, compared with MMCD, SQNGD replaces sequential coordinate-wise discrete search with batched FFT-based gradient-based updates. This makes SQNGD more suitable for parallelization  and GPU-accelerated implementation, so its runtime advantage is expected to become more evident as the problem size increases.

\section{NUMERICAL RESULTS AND ANALYSIS}
In this section, numerical experiments are presented to evaluate the performance of the proposed SQNGD in four aspects. First, SQNGD is compared with SQN and MMCD in terms of correlation properties, in order to demonstrate that the joint transmit--receive design can also improve the correlation performance. It is worth noting that, when the Doppler variable is  $f=0$, AF calculation is essentially a correlation operation. Second, the effects of several key parameters, including the sequence length $N$, phase alphabet size $B$, SNRL $\tilde{\mu}$, and Doppler range $f$, are investigated. Third, the delay--Doppler sidelobe performance of SQNGD is compared with that of an AF extension of SQN \cite{11079603} and MMCD \cite{10679908}. Finally, the optimization efficiency of SQNGD is evaluated in terms of computational time and empirical convergence behavior. The AF extension of SQN, termed SQNAF, is obtained by replacing the correlation-based PSL objective in SQN with the delay--Doppler WPSL objective while no independent receive-filter optimization is introduced. The network architecture and hyperparameter settings are kept identical to those used in SQNGD and SQN for a fair comparison. Among them, MMCD is also a joint transmit--receive design method that employs CD to optimize the discrete transmit sequences and MM to optimize the corresponding receive filters while considering the delay--Doppler ROI, and is therefore selected as the main baseline.

To evaluate the PSL performance using a normalized metric, the Peak Sidelobe Ratio (PSLR) is defined as
\begin{equation}
\operatorname{PSLR} ( \text{dB}) = 20 \log_{10} \frac{PSL}{N}.
\label{eq:PSLR}
\end{equation}
The experiments are conducted on a workstation equipped with an Intel Ultra 9 275HX CPU and an NVIDIA RTX 5080 Laptop GPU. In the implementation, SQN, SQNAF and SQNGD are accelerated by GPU, whereas MMCD is executed on CPU, since MMCD is primarily designed as a CPU algorithm. For the implementation of SQNGD, the transmitter-side network adopts a ResNet architecture with 5 residual blocks and a hidden width of 128, while the hidden and residual layers mainly use the tanh activation function, and a sigmoid activation is adopted at the output layer. The Adam learning rates are set to $\eta_h=5\times10^{-2}$ and $\eta_x=1\times10^{-5}$ for receive-filter and transmit-waveform updates, respectively. The inner iteration numbers are chosen as $N_h=10$ and $N_x=10$, and the outer iteration number is set to $T=2000$. The loss tradeoff parameter is fixed as $\varepsilon=0.1$ in the reported experiments. In addition, the copying factor is set to $r_n=100$, and the steepness parameter is chosen as $c=300$.

\subsection{Comparisons of Auto- and Cross- Correlations  Among SQN, SQNGD, and MMCD}

In this subsection, the correlation properties of SQN, SQNGD, and MMCD are compared. Since SQN adopts the transmit-only design with matched filter, the target SNRL is set to $\tilde{\mu}=0$~dB to ensure a fair comparison, and the number of transmit antennas is fixed as $M=2$. The sequence length $N$ is set to $512$ or $1200$, and the phase alphabet size $B$ is set to $4$ or $16$. The ROI is defined as $\tau \in [-N+1,\,N-1]$.
Although SQNGD is developed for suppressing delay--Doppler sidelobes of the AF, it is still meaningful to examine the correlation performance of the resulting transmit waveforms. Such a comparison helps reveal whether the proposed joint transmit--receive design can preserve competitive correlation properties while extending the optimization objective from correlation sidelobe suppression to delay--Doppler sidelobe suppression.

\begin{figure*}[!t]
\centering
\includegraphics[width=\textwidth]{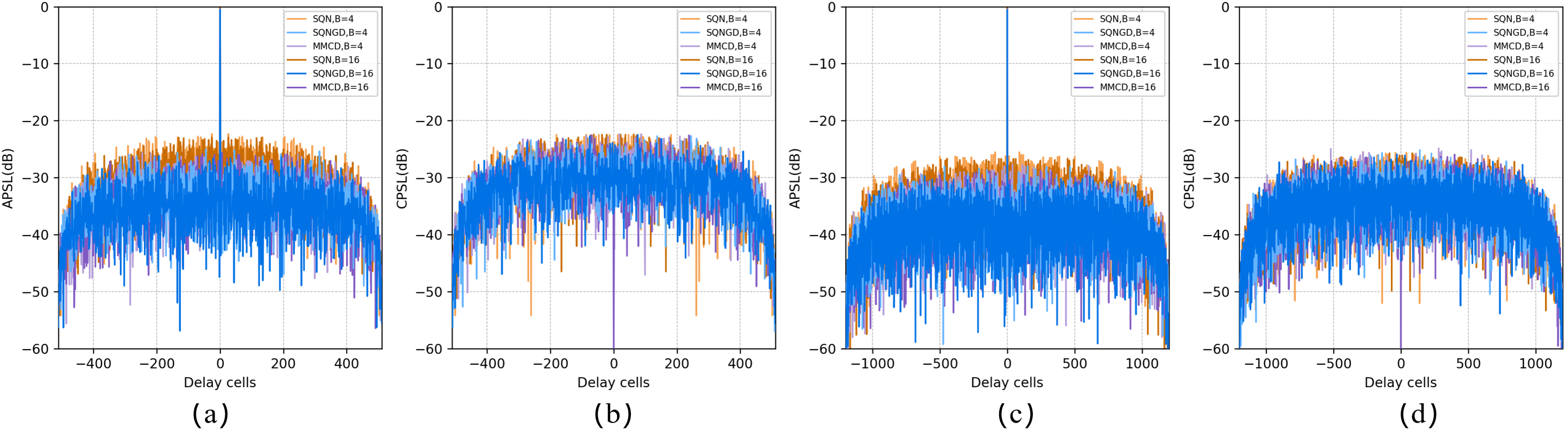}
\caption{Comparison of the correlation results of SQN, SQNGD, and MMCD for different sequence length $N$ and phase alphabet sizes $B$. 
(a) Auto-correlation comparison for $N=512$; 
(b) Cross-correlation comparison for $N=512$; 
(c) Auto-correlation comparison for $N=1200$; 
(d) Cross-correlation comparison for $N=1200$.}
\label{fig:corr_compare_all}
\end{figure*}

As shown in Table~\ref{tab:correlation_properties_m2}, SQNGD achieves the best APSL among the three methods under all tested settings. Compared with SQN, the APSL improvement of SQNGD is about $2.45$--$3.30$~dB. In terms of CPSL, SQNGD remains competitive in most settings. Specifically, SQNGD slightly outperforms SQN for $N=512$ and $B=4$, $N=512$ and $B=16$, and $N=1200$ and $B=16$, while a small degradation of about $0.32$~dB is observed when $N=1200$ and $B=4$. Compared with MMCD, SQNGD achieves lower APSL in most cases, while the CPSL values of the two methods are generally close.
These results indicate that extending the soft-quantization idea to the joint transmit--receive setting does not sacrifice the correlation quality of the designed waveforms. On the contrary, SQNGD consistently provides stronger correlation APSL suppression and maintains competitive CPSL performance. The corresponding auto- and cross-correlation results are further illustrated in Fig.~\ref{fig:corr_compare_all}, where SQNGD generally exhibits lower sidelobe levels than SQN and comparable performance to MMCD under different sequence lengths and phase alphabet sizes.
\begin{table}[!t]
\centering
\caption{Comparison of Correlation APSL and CPSL (dB) Among SQN, SQNGD, and MMCD for $M=2$, $\tilde{\mu}=0$~dB}
\label{tab:correlation_properties_m2}
\resizebox{\columnwidth}{!}{%
\begin{tabular}{cc|ccc|ccc}
\hline
\multicolumn{1}{c}{$N$} &
\multicolumn{1}{c}{$B$} &
\multicolumn{3}{c|}{\textbf{APSL (dB)}} &
\multicolumn{3}{c}{\textbf{CPSL (dB)}} \\
\cline{3-8}
 &  & \multicolumn{1}{c}{SQN} & \multicolumn{1}{c}{SQNGD} & \multicolumn{1}{c|}{MMCD}
    & \multicolumn{1}{c}{SQN} & \multicolumn{1}{c}{SQNGD} & \multicolumn{1}{c}{MMCD} \\
\hline
$512$  & $4$  & $-22.32$ & $\textbf{-25.03}$ & $-25.63$ & $-22.44$ & $\textbf{-22.66}$ & $-22.34$ \\
$512$  & $16$ & $-22.92$ & $\textbf{-26.22}$ & $-25.61$ & $-22.39$ & $\textbf{-22.48}$ & $-22.54$ \\
$1200$ & $4$  & $-25.52$ & $\textbf{-27.97}$ & $-27.66$ & $-25.47$ & $\textbf{-25.15}$ & $-24.84$ \\
$1200$ & $16$ & $-26.64$ & $\textbf{-29.71}$ & $-28.76$ & $-25.69$ & $\textbf{-25.88}$ & $-25.94$ \\
\hline
\end{tabular}%
}
\end{table}

\subsection{The Effect of Diverse Waveform Parameters and Ranges of Interested Doppler}
In this subsection, the effects of several key parameters, including the sequence length $N$, phase alphabet size $B$, SNRL $\tilde{\mu}$, and Doppler range $f$, on sidelobe suppression performance are investigated. The analysis focuses on how these parameters influence the PSL under different settings. In all experiments, unless otherwise specified, the number of antennas is set to $M=2$, the ROI is defined as $\tau \in [-50,\,50]$, the Doppler range is set to $f \in [-0.5,\,0.5]$, the sequence length and phase alphabet size are fixed as $N=512$ and $B=4$, and the target SNRL is set to $\tilde{\mu}=0.5$~dB, respectively.

\begin{figure}[!t]
\centering
\includegraphics[width=\linewidth]{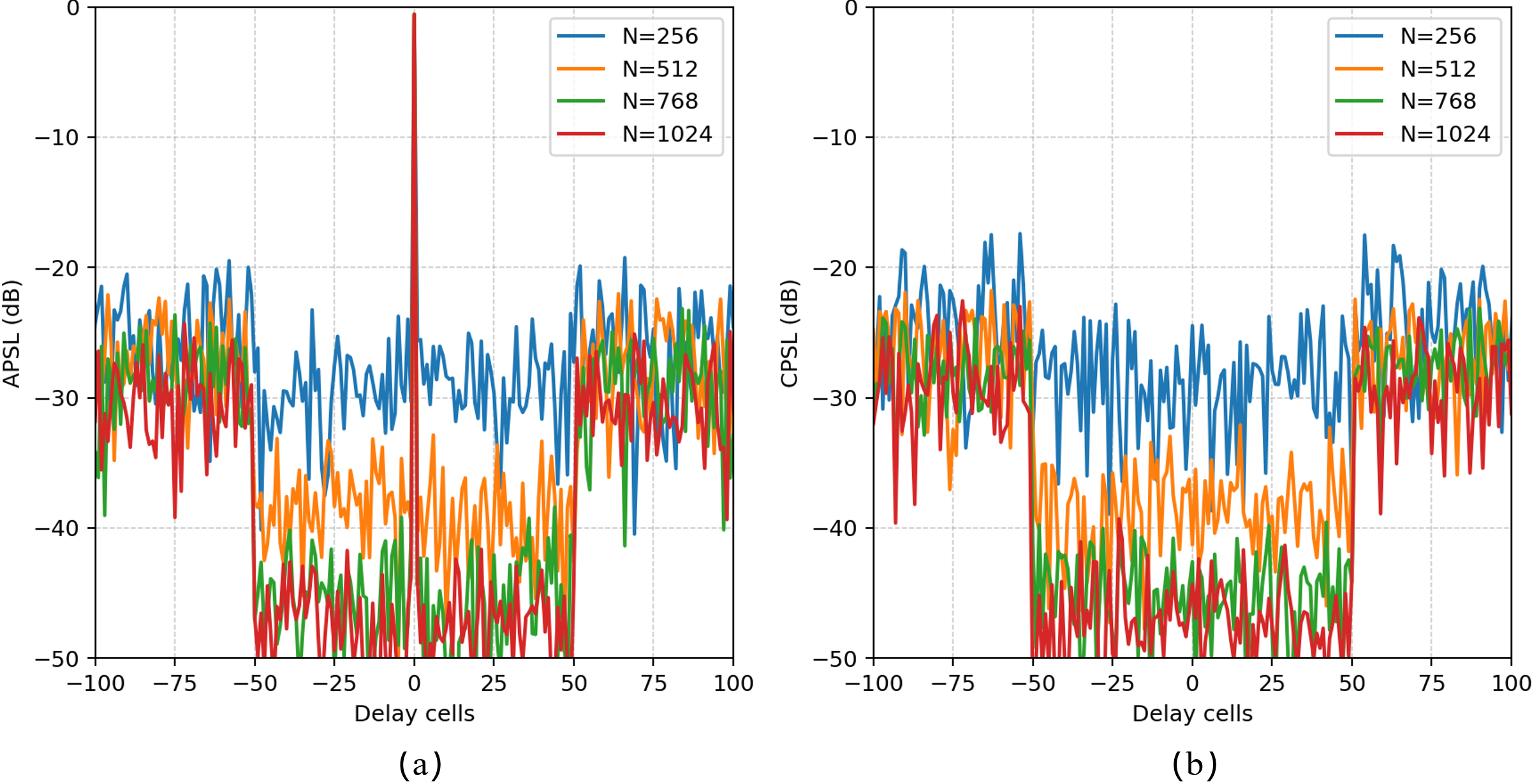}
\caption{Effect of different sequence lengths $N$ on AF and CAF performances for $M=2$, $B=4$, $f \in [-0.5,\,0.5]$ and $\tilde{\mu}=0.5$~dB. (a) Zero-Doppler AF slice. (b) Zero-Doppler CAF slice.}
\label{fig:effect_of_N_m2_b4}
\end{figure}

\textbf{Sequence Length $N$:}
Fig.~\ref{fig:effect_of_N_m2_b4} illustrates the effect of different sequence lengths $N$ on the AF and CAF PSL performance for $M=2$ and $B=4$. The optimization is carried out over the Doppler range $f \in [-0.5,\,0.5]$, whereas the figure shows the slice at $f=0$. It can be observed that, as $N$ increases from $256$ to $1024$, the sidelobe levels within the ROI are generally reduced in both the AF and CAF cases, indicating that a longer sequence provides more design degrees of freedom for sidelobe suppression. Meanwhile, the difference outside the optimized delay interval is less pronounced, which is consistent with the fact that the optimization mainly emphasizes the selected ROI. These results show that increasing the sequence length is beneficial for improving the sidelobe suppression performance of the designed waveforms.

\begin{figure}[!t]
\centering
\includegraphics[width=\linewidth]{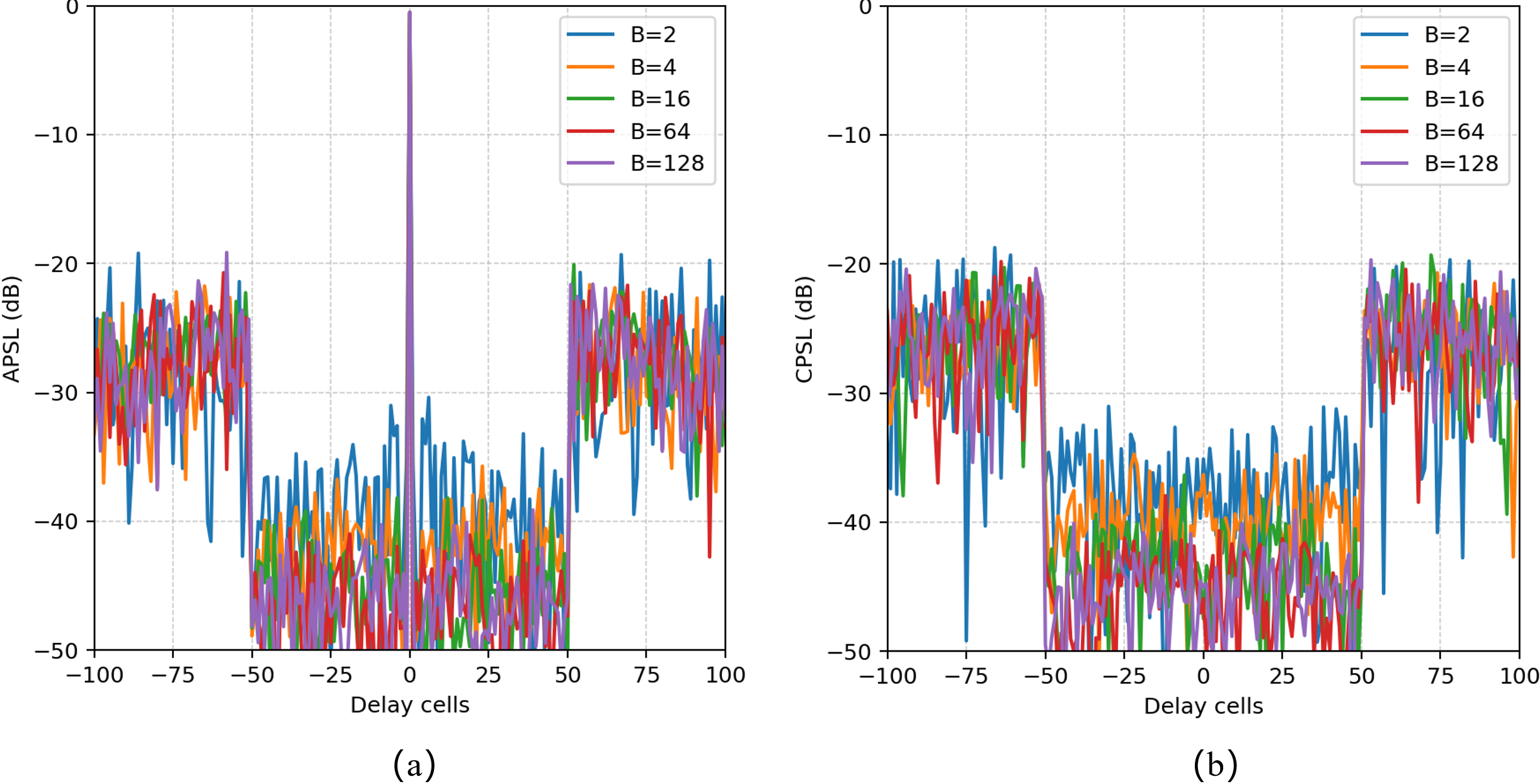}
\caption{Effect of phase alphabet sizes $B$ on the AF and CAF performances for $M=2$, $N=512$, $f \in [-0.5,\,0.5]$ and $\tilde{\mu}=0.5$~dB.  (a) Zero-Doppler AF slice. (b) Zero-Doppler CAF slice.}
\label{fig:effect_of_B_m2_n512}
\end{figure}

\textbf{Phase Alphabet Size $B$:}
Fig.~\ref{fig:effect_of_B_m2_n512} shows the effect of different phase alphabet sizes $B$ on the AF and CAF PSL performance for $M=2$ and $N=512$. The optimization is carried out over the Doppler range $f \in [-0.5,\,0.5]$, whereas the figure shows the slice at $f=0$. As the phase alphabet becomes richer, the sidelobes within the ROI are gradually suppressed. The binary case with $B=2$ exhibits the highest sidelobe level, while larger phase alphabet sizes, such as $B=16$, $64$, and $128$, achieve lower and more stable sidelobe floors. This behavior indicates that increasing $B$ provides greater flexibility in DPW design and makes it easier to approach a more favorable sidelobe pattern. At the same time, the performance gap becomes smaller when $B$ is already sufficiently large, which suggests diminishing returns beyond a certain phase alphabet size.

\textbf{SNRL $\tilde{\mu}$:}
Table~\ref{tab:snrl_compare_sqnaf_sqngd_mmcd_m2_n512_b4} compares the delay--Doppler sidelobe performance of SQNAF, SQNGD, and MMCD under different SNRL settings for $M=2$, $N=512$, and $B=4$. Since SQNAF does not involve independent receive-filter optimization, only the result at $\tilde{\mu}=0$~dB is reported for SQNAF. It can be observed that SQNAF yields the worst sidelobe performance at $\tilde{\mu}=0$~dB, since it does not employ joint transmit--receive optimization. For SQNGD and MMCD, it can be observed that the APSL and CPSL improve as the allowable SNRL $\tilde{\mu}$ increases from $0$~dB to $1.5$~dB. This is because a looser SNRL constraint provides greater flexibility for receive-filter optimization, thereby allowing a better suppression of sidelobes at the cost of a limited mainlobe loss. When $\tilde{\mu}=0$~dB, SQNGD is already slightly better than MMCD in both APSL and CPSL. As the SNRL constraint is gradually relaxed, the advantage of SQNGD becomes more evident. In particular, at $\tilde{\mu}=0.5$~dB, SQNGD improves APSL and CPSL by about $3.18$~dB and $4.27$~dB, respectively, over MMCD. When $\tilde{\mu}=1$~dB, the corresponding improvements become about $4.50$~dB and $3.43$~dB. When $\tilde{\mu}=1.5$~dB, SQNGD still achieves about $5.27$~dB and $4.08$~dB lower APSL and CPSL than MMCD, respectively. These results indicate that SQNGD can exploit the increased design freedom under relaxed SNRL constraints more effectively than MMCD.

\textbf{Range of Interested Doppler $f$:}
Table~\ref{tab:fd_compare_sqnaf_sqngd_mmcd_m2_n1200_b16} reports the APSL and CPSL performance of SQNAF, SQNGD, and MMCD under different Doppler ranges for $M=2$, $N=1200$, and $B=16$. Since the Doppler variable is defined over $f \in [-N/2,\,N/2]$, different Doppler ranges from $[-0.1,\,0.1]$ to $[-600,\,600]$ are considered in the experiments. It can be observed that, as the Doppler range expands from $[-0.1,\,0.1]$ to $[-600,\,600]$, the APSL and CPSL performance of all three methods gradually degrades. This indicates that the sidelobe suppression problem becomes more challenging when a broader delay--Doppler region must be controlled simultaneously. Nevertheless, SQNGD consistently outperforms both SQNAF and MMCD over all tested Doppler ranges. Even for the narrowest range $[-0.1,\,0.1]$, SQNGD already achieves clear improvements, with about $23.73$~dB and $24.92$~dB lower APSL and CPSL than SQNAF, respectively, and about $6.71$~dB and $8.20$~dB lower APSL and CPSL than MMCD. As the Doppler range becomes wider, the advantage of SQNGD remains stable. For example, over the range $[-0.5,\,0.5]$, SQNGD improves APSL and CPSL by about $18.36$~dB and $20.10$~dB over SQNAF, respectively, and by about $3.52$~dB and $5.85$~dB over MMCD. Even over the widest range $[-600,\,600]$, the corresponding improvements of SQNGD remain about $6.72$~dB and $9.56$~dB over SQNAF, and about $3.22$~dB and $3.45$~dB over MMCD. It is also observed that SQNAF performs significantly worse than both SQNGD and MMCD over all tested Doppler ranges, which indicates that simply extending SQN from correlation optimization to AF optimization is insufficient. The substantial gain mainly comes from the additional design freedom provided by joint transmit--receive optimization. These results demonstrate that SQNGD is more effective than both SQNAF and MMCD for delay--Doppler sidelobe suppression over different Doppler ranges.

Overall, under all these settings, SQNGD exhibits stable and competitive performance, which further verifies its effectiveness for joint transmit--receive design of Doppler-resilient DPWs.

\begin{table}[!t]
\centering
\caption{Comparison of APSL and CPSL (dB) Among SQNAF, SQNGD, and MMCD Under Different SNRL Settings for $M=2$, $N=512$, $B=4$, and $f \in [-0.5,\,0.5]$}
\label{tab:snrl_compare_sqnaf_sqngd_mmcd_m2_n512_b4}
\resizebox{\columnwidth}{!}{%
\begin{tabular}{c|ccc|ccc}
\hline
\multirow{2}{*}{\textbf{SNRL $\tilde{\mu}$ (dB)}} 
& \multicolumn{3}{c|}{\textbf{APSL (dB)}} 
& \multicolumn{3}{c}{\textbf{CPSL (dB)}} \\
\cline{2-7}
& \textbf{SQNAF} & \textbf{SQNGD} & \textbf{MMCD}
& \textbf{SQNAF} & \textbf{SQNGD} & \textbf{MMCD} \\
\hline
$0$   & $-22.39$ & $\mathbf{-24.10}$ & $-24.01$ & $-20.83$ & $\mathbf{-24.31}$ & $-23.83$ \\
$0.5$ & N/A      & $\mathbf{-29.63}$ & $-26.45$ & N/A      & $\mathbf{-29.07}$ & $-24.80$ \\
$1$   & N/A      & $\mathbf{-32.32}$ & $-27.82$ & N/A      & $\mathbf{-30.89}$ & $-27.46$ \\
$1.5$ & N/A      & $\mathbf{-33.11}$ & $-27.84$ & N/A      & $\mathbf{-31.74}$ & $-27.66$ \\
\hline
\end{tabular}%
}
\end{table}

\begin{table}[!t]
\centering
\caption{Comparison of APSL and CPSL (dB) Among SQNAF, SQNGD, and MMCD Under Different Doppler Ranges for $M=2$, $N=1200$, $B=16$, and $\tilde{\mu}=0.5$~dB}
\label{tab:fd_compare_sqnaf_sqngd_mmcd_m2_n1200_b16}
\resizebox{\columnwidth}{!}{%
\begin{tabular}{c|ccc|ccc}
\hline
\textbf{$f$ range} & \multicolumn{3}{c|}{\textbf{APSL (dB)}} & \multicolumn{3}{c}{\textbf{CPSL (dB)}} \\
\cline{2-7}
 & \textbf{SQNAF} & \textbf{SQNGD} & \textbf{MMCD} & \textbf{SQNAF} & \textbf{SQNGD} & \textbf{MMCD} \\
\hline
$[-0.1,\,0.1]$   & $-25.14$ & $\mathbf{-48.87}$ & $-42.16$ & $-24.18$ & $\mathbf{-49.10}$ & $-40.90$ \\
$[-0.5,\,0.5]$   & $-24.73$ & $\mathbf{-43.09}$ & $-39.57$ & $-23.55$ & $\mathbf{-43.65}$ & $-37.80$ \\
$[-1,\,1]$       & $-24.54$ & $\mathbf{-41.49}$ & $-37.43$ & $-23.25$ & $\mathbf{-41.09}$ & $-37.05$ \\
$[-3,\,3]$       & $-24.04$ & $\mathbf{-36.10}$ & $-33.19$ & $-22.41$ & $\mathbf{-34.27}$ & $-32.38$ \\
$[-10,\,10]$     & $-23.92$ & $\mathbf{-34.06}$ & $-30.08$ & $-21.63$ & $\mathbf{-33.45}$ & $-30.33$ \\
$[-20,\,20]$     & $-23.94$ & $\mathbf{-33.65}$ & $-29.57$ & $-21.34$ & $\mathbf{-33.19}$ & $-29.57$ \\
$[-600,\,600]$   & $-24.66$ & $\mathbf{-31.38}$ & $-28.16$ & $-21.69$ & $\mathbf{-31.25}$ & $-27.80$ \\
\hline
\end{tabular}%
}
\end{table}
\subsection{ Comparisons of AFs and CAFs Among SQNAF, SQNGD, and MMCD}
\begin{table}[!t]
\centering
\caption{Comparison of APSL and CPSL (dB) Among SQNAF, SQNGD, and MMCD for $M=2$,  $f \in [-0.5,\,0.5]$, and $\tilde{\mu}=0.5$~dB}
\label{tab:apsl_cpsl_sqn_sqngd_mmcd_m=2}
\resizebox{\columnwidth}{!}{%
\begin{tabular}{cc|ccc|ccc}
\hline
\multicolumn{1}{c}{$N$} &
\multicolumn{1}{c}{$B$} &
\multicolumn{3}{c|}{\textbf{APSL (dB)}} &
\multicolumn{3}{c}{\textbf{CPSL (dB)}} \\
\cline{3-8}
 &  & \multicolumn{1}{c}{SQNAF} & \multicolumn{1}{c}{SQNGD} & \multicolumn{1}{c|}{MMCD}
    & \multicolumn{1}{c}{SQNAF} & \multicolumn{1}{c}{SQNGD} & \multicolumn{1}{c}{MMCD} \\
\hline
$512$  & $4$  & $-22.39$ & $\textbf{-29.63}$ & $-26.45$ & $-20.83$ & $\textbf{-29.07}$ & $-24.80$ \\
$512$  & $16$ & $-22.62$ & $\textbf{-33.36}$ & $-29.74$ & $-20.14$ & $\textbf{-30.17}$ & $-29.25$ \\
$1200$ & $4$  & $-25.17$ & $\textbf{-37.69}$ & $-33.15$ & $-23.53$ & $\textbf{-38.93}$ & $-32.57$ \\
$1200$ & $16$ & $-24.73$ & $\textbf{-43.09}$ & $-39.57$ & $-23.55$ & $\textbf{-43.65}$ & $-37.80$ \\
\hline
\end{tabular}%
}
\end{table}

\begin{table}[!t]
\centering
\caption{Comparison of APSL and CPSL (dB) Among SQNAF, SQNGD, and MMCD for $M=10$,  $f \in [-0.5,\,0.5]$, and $\tilde{\mu}=0.5$~dB}
\label{tab:apsl_cpsl_sqn_sqngd_mmcd_m=10}
\resizebox{\columnwidth}{!}{%
\begin{tabular}{cc|ccc|ccc}
\hline
\multicolumn{1}{c}{$N$} &
\multicolumn{1}{c}{$B$} &
\multicolumn{3}{c|}{\textbf{APSL (dB)}} &
\multicolumn{3}{c}{\textbf{CPSL (dB)}} \\
\cline{3-8}
 &  & \multicolumn{1}{c}{SQNAF} & \multicolumn{1}{c}{SQNGD} & \multicolumn{1}{c|}{MMCD}
    & \multicolumn{1}{c}{SQNAF} & \multicolumn{1}{c}{SQNGD} & \multicolumn{1}{c}{MMCD} \\
\hline
$512$  & $4$  & $-21.2$ & $\textbf{-21.02}$ & $-18.56$ & $-17.2$ & $\textbf{-18.76}$ & $-16.43$ \\
$512$  & $16$ & $-20.44$ & $\textbf{-22.24}$ & $-20.96$ & $-17.62$ & $\textbf{-19.26}$ & $-17.96$ \\
$1200$ & $4$  & $-23.53$ & $\textbf{-28.43}$ & $-25.36$ & $-20.58$ & $\textbf{-25.18}$ & $-24.12$ \\
$1200$ & $16$ & $-23.33$ & $\textbf{-30.71}$ & $-28.41$ & $-20.84$ & $\textbf{-30.62}$ & $-25.81$ \\
\hline
\end{tabular}%
}
\end{table}

\begin{figure*}[!t]
  \centering
  \includegraphics[width=\textwidth]{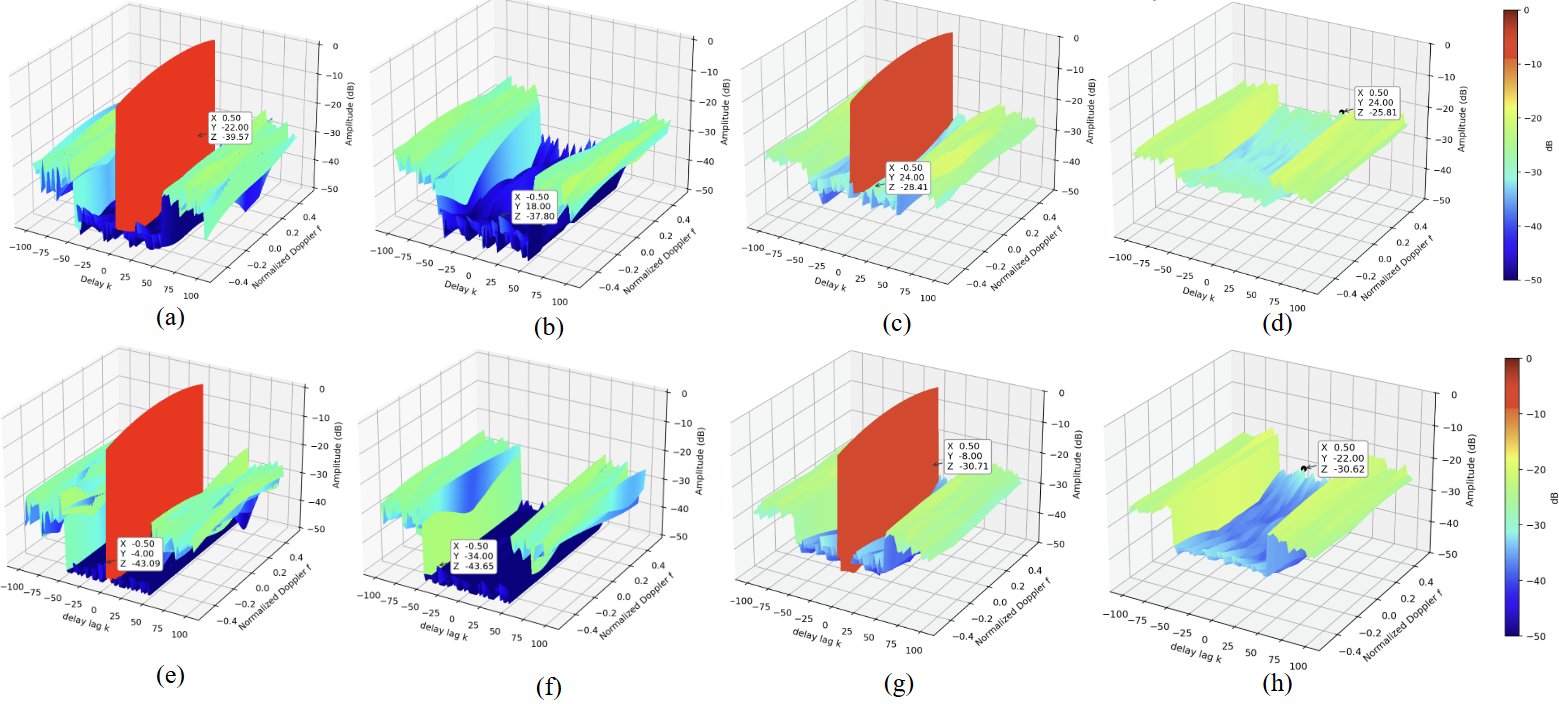}
\caption{Comparisons of the AFs and CAFs of MMCD and SQNGD for $N=1200$, $B=16$, $f \in [-0.5,\,0.5]$, and $\tilde{\mu}=0.5$~dB. (a) AF of MMCD for $M=2$; (b) CAF of MMCD for $M=2$; (c) AF of MMCD for $M=10$; (d) CAF of MMCD for $M=10$; (e) AF of SQNGD for $M=2$; (f) CAF of SQNGD for $M=2$; (g) AF of SQNGD for $M=10$; (h) CAF of SQNGD for $M=10$.}
  \label{fig:AF_compare}
\end{figure*}

In this subsection, the delay--Doppler sidelobe performance of SQNGD is compared with SQNAF and MMCD. Two multi-antenna scenarios, namely $M=2$ and $M=10$, are considered to evaluate the performance of SQNGD under different multi-antenna settings. The normalized Doppler frequency is set to $f \in [-0.5,\,0.5]$, and the delay ROI is fixed to $\tau \in [-50,\,50]$. This fixed ROI is adopted to facilitate a clear comparison across different sequence lengths and different phase alphabet sizes. Meanwhile, the target SNRL of SQNGD and MMCD is fixed at $\tilde{\mu}=0.5$~dB. Under these settings, the PSL performance of SQNAF, SQNGD, and MMCD is evaluated for two sequence lengths, namely $N=512$ and $N=1200$, and two phase alphabet sizes, namely $B=4$ and $B=16$.

Table~\ref{tab:apsl_cpsl_sqn_sqngd_mmcd_m=2} and Table~\ref{tab:apsl_cpsl_sqn_sqngd_mmcd_m=10} summarize the APSL and CPSL performance of SQNAF, SQNGD, and MMCD for $M=2$ and $M=10$, respectively. Compared with MMCD, SQNGD consistently achieves lower sidelobe levels across all tested configurations, demonstrating stronger sidelobe suppression capability. Quantitatively, the APSL gains of SQNGD over MMCD are $3.18$--$4.54$~dB for $M=2$ and $1.28$--$3.07$~dB for $M=10$, respectively. SQNGD also provides clear CPSL advantages, with improvements of $0.92$--$6.36$~dB for $M=2$ and $1.06$--$4.81$~dB for $M=10$. When compared with the SQNAF, SQNGD also achieves lower sidelobe levels in most cases. For $M=2$, the APSL and CPSL improvements reach $7.24$--$18.36$~dB and $8.24$--$20.10$~dB, respectively. For $M=10$, SQNGD remains better than SQNAF in most configurations, with APSL and CPSL improvements of up to $7.38$~dB and $9.78$~dB, respectively, except that the APSL is slightly worse by about $0.18$~dB when $N=512 $ and $ B=4$. It is also observed that SQNAF performs significantly worse than both SQNGD and MMCD in most cases. This is mainly because SQN only extends the optimization objective to the AF domain, but does not introduce joint transmit--receive design. As a result, the additional design freedom provided by receive-filter optimization cannot be exploited, leading to inferior sidelobe suppression performance. 

It is further observed that the APSL and CPSL improvements for $M=10$ are generally smaller than those for $M=2$. This is because increasing $M$ substantially enlarges the set of AFs that need to be jointly optimized: the number of auto-AF terms scales linearly with $M$, and thus increases by a factor of $10/2=5$ when $M$ increases from $2$ to $10$. More importantly, the number of cross-AF terms grows combinatorially as $C_M^2$, where $C_l^k$ denotes the binomial coefficient. Hence the number of CAFs increases by a factor of $C_{10}^2 / C_2^2 = 45$ when $M$ increases from $2$ to $10$. As a result, the optimization becomes more constrained and the achievable sidelobe suppression margin tends to decrease. Nevertheless, SQNGD still yields consistent improvements over MMCD for $M=10$ in all cases, indicating that SQNGD remains effective even under substantially increased multiwaveform coupling.

These numerical results are also visually confirmed by Fig.~\ref{fig:AF_compare}, which presents the AF and CAF surfaces of MMCD and SQNGD for the representative setting with $N=1200$ and $B=16$, under $M=2$ and $M=10$. It can be observed that the AF and CAF surfaces generated by SQNGD exhibit lower sidelobe floors and cleaner local structures within the prescribed delay--Doppler ROI. In particular, compared with MMCD, SQNGD produces more pronounced sidelobe suppression in both AF and CAF, which is consistent with the APSL and CPSL gains reported in Table~\ref{tab:apsl_cpsl_sqn_sqngd_mmcd_m=2} and Table~\ref{tab:apsl_cpsl_sqn_sqngd_mmcd_m=10}.

\subsection{Optimization Efficiency and Empirical Convergence Behavior}

\begin{table}[!t]
\centering
\caption{Optimization Time Comparison (seconds) for $f \in [-0.5,\,0.5]$ and $\tilde{\mu}=0.5$~dB}
\label{tab:time_sqnaf_sqngd_mmcd}
\resizebox{\columnwidth}{!}{%
\begin{tabular}{cc|ccc|ccc}
\hline
\multicolumn{1}{c}{$N$} &
\multicolumn{1}{c}{$B$} &
\multicolumn{3}{c|}{\textbf{$M=2$}} &
\multicolumn{3}{c}{\textbf{$M=10$}} \\
\cline{3-8}
 &  & \multicolumn{1}{c}{SQNAF} & \multicolumn{1}{c}{SQNGD} & \multicolumn{1}{c|}{MMCD}
    & \multicolumn{1}{c}{SQNAF} & \multicolumn{1}{c}{SQNGD} & \multicolumn{1}{c}{MMCD} \\
\hline
$512$  & $4$  & $49.44$ & $\mathbf{34.20}$ & $113.10$ & $25.03$ & $\mathbf{45.60}$ & $338.91$ \\
$512$  & $16$ & $82.32$ & $\mathbf{116.10}$ & $217.84$ & $41.32$ & $\mathbf{128.60}$ & $483.29$ \\
$1200$ & $4$  & $32.56$ & $\mathbf{36.50}$ & $182.54$ & $43.72$ & $\mathbf{58.40}$ & $640.63$ \\
$1200$ & $16$ & $84.53$ & $\mathbf{148.30}$ & $399.78$ & $44.72$ & $\mathbf{158.90}$ & $802.98$ \\
\hline
\end{tabular}%
}
\end{table}
\begin{figure}[!t]
  \centering
  \includegraphics[width=\columnwidth]{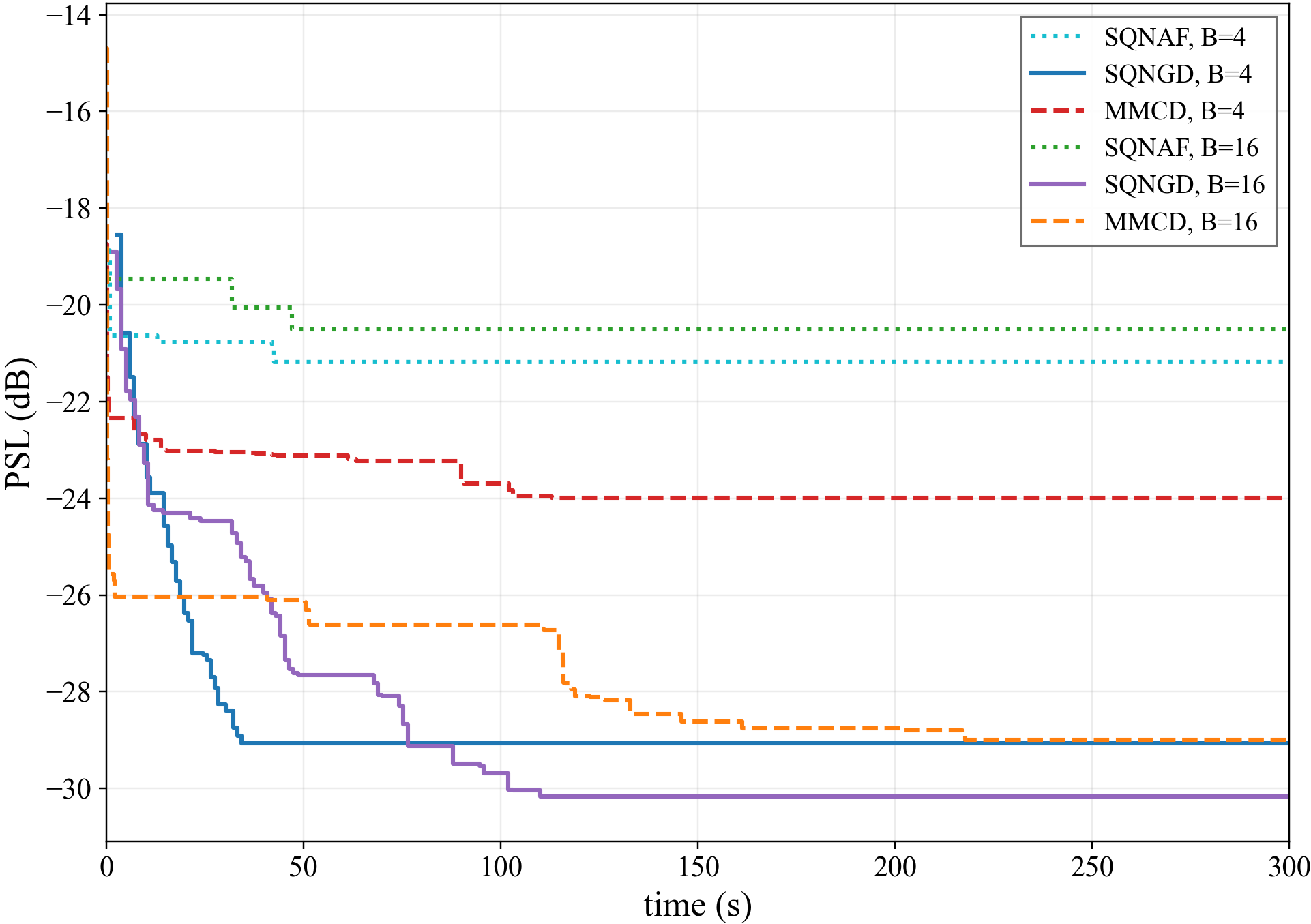}
  \caption{The PSL along with the computational time for SQNGD, SQNAF, and MMCD with $M=2$, $N=512$, $B=4$ and $B=16$, $f \in [-0.5,\,0.5]$, and $\tilde{\mu}=0.5$~dB.}
  \label{fig:convergence_psl}
\end{figure}

The optimization efficiency is an important metric for large-scale DPW design. Conventional BCD-based methods typically rely on sequential block updates and local discrete searches, which lead to relatively high computational cost and limited scalability as the waveform dimension increases. By contrast, SQNGD adopts gradient-based alternating optimization together with FFT-accelerated AF evaluation, thereby avoiding exhaustive coordinate-wise discrete searches and achieving better computational efficiency in practice.

Table~\ref{tab:time_sqnaf_sqngd_mmcd} summarizes the overall optimization time of SQNAF, SQNGD, and MMCD under different parameter settings. It can be observed that SQNGD is consistently faster than MMCD in all listed cases, and the advantage becomes more pronounced as the number of transmit antennas $M$ increases. For $M=2$, SQNGD reduces the runtime from $113.10$--$399.78$~s to $34.20$--$148.30$~s, corresponding to about $1.9\times$--$4.9\times$ speed-up. For $M=10$, the runtime reduction is even more significant: SQNGD completes in $45.60$--$158.90$~s, whereas MMCD requires $338.91$--$802.98$~s, yielding about $3.7\times$--$11.0\times$ acceleration. Compared with SQNAF, SQNGD generally requires more optimization time in most cases, since SQNGD further introduces joint receive-filter optimization. Nevertheless, this additional computational cost brings clear gains in delay--Doppler sidelobe suppression. For example, for $M=2$, $N=1200$, and $B=16$, the runtime increases from $84.53$~s for SQNAF to $148.30$~s for SQNGD, while the APSL and CPSL are improved from $-24.73$~dB and $-23.55$~dB to $-43.09$~dB and $-43.65$~dB, respectively. Similarly, for $M=10$, $N=1200$, and $B=16$, the runtime increases from $44.72$~s to $158.90$~s, while the APSL and CPSL are improved from $-23.33$~dB and $-20.84$~dB to $-30.71$~dB and $-30.62$~dB, respectively. These results indicate that the additional runtime of SQNGD is justified by the substantial performance gain brought by joint transmit--receive optimization.

To further illustrate the empirical convergence behavior, Fig.~\ref{fig:convergence_psl} plots the PSL evolution with computational time for a representative case with $M=2$, $N=512$, under two phase alphabet sizes, namely $B=4$ and $B=16$. For both values of $B$, MMCD reduces the PSL slightly faster during the very early stage. However, as the optimization proceeds, the improvement of MMCD gradually slows down, whereas SQNGD continues to reduce the PSL more effectively and eventually achieves lower final sidelobe levels. Compared with SQNAF, SQNGD also achieves substantially lower final PSL values. Although SQNGD requires longer optimization time than SQNAF, it provides about $8$--$10$~dB lower final PSL in the representative cases shown in Fig.~\ref{fig:convergence_psl}, which further confirms the benefit of introducing joint transmit--receive optimization rather than relying only on transmitter-side AF optimization. These observations are consistent with the runtime comparison reported in Table~\ref{tab:time_sqnaf_sqngd_mmcd}, further confirming that SQNGD not only converges faster than MMCD but also provides better final sidelobe performance under the same representative settings.

Overall, the results in Table~\ref{tab:time_sqnaf_sqngd_mmcd} and Fig.~\ref{fig:convergence_psl} demonstrate that SQNGD provides both higher optimization efficiency and more favorable empirical convergence behavior than MMCD. Although SQNGD generally requires more runtime than SQNAF, this additional computational cost yields a substantial improvement in delay--Doppler sidelobe suppression, further demonstrating the effectiveness of joint transmit--receive design for Doppler-resilient DPW optimization.

\section{CONCLUSIONS}
 This paper proposed SQNGD, a joint framework for designing Doppler-resilient unimodular DPWs and receive filters. A MOOP was formulated to achieve the Pareto trade-off among AF, CAF, and SNRL. To tackle this nonconvex NP-hard problem, soft-to-hard quantization and gradient descent were employed to alternately optimize the DPWs under unimodular constraints and the receive filters,  respectively. To improve computational efficiency, an FFT-accelerated model is used in alternating optimization process. Numerical results demonstrated the effectiveness and efficiency of SQNGD. It consistently outperformed the SOTA schemes, namely MMCD and SQN, in both delay–Doppler sidelobe suppression and convergence speed.  In particular, for $M=2$, $N=1200$, and $B=16$ with ROI $f \in [-0.5,\,0.5]$ and $\tau \in [-50,\,50]$, SQNGD achieved APSL 
of $-43.09$~dB and  CPSL of $-43.65$~dB, respectively. Compared with MMCD, the improvements were $3.52$~dB in APSL and $5.85$~dB in CPSL, respectively. Compared with SQNAF, the gains reached $18.36$~dB in APSL and $20.10$~dB in CPSL, respectively. Moreover, SQNGD attained approximately  $1.9\times$ to $11.0\times$ speed-up over the MMCD across all experimental settings.


 



\bibliographystyle{IEEEtran}  %
\bibliography{references}     %

@ARTICLE{10128881,
  author={Zhong, Kai and Hu, Jinfeng and Pan, Cunhua},
  journal={IEEE Transactions on Instrumentation and Measurement}, 
  title={Constant Modulus {MIMO} Radar Waveform Design via Iterative Optimization Network Method}, 
  year={2023},
  volume={72},
  number={},
  pages={1-11},
  doi={10.1109/TIM.2023.3277111}}

@ARTICLE{10348020,
  author={Eamaz, Arian and Yeganegi, Farhang and Soltanalian, Mojtaba},
  journal={IEEE Transactions on Signal Processing}, 
  title={MaRLI: Attack on the Discrete-Phase {W{ISL}} Minimization Problem}, 
  year={2024},
  volume={72},
  number={},
  pages={219-234},
  doi={10.1109/TSP.2023.3339398}}

@ARTICLE{10620365,
  author={Zhong, Kai and Hu, Jinfeng and Yuan, Ye and Wang, Yuankai and Teh, Kah Chan and Pan, Cunhua and Li, Huiyong and Yu, Xianxiang and Cui, Guolong},
  journal={IEEE Transactions on Cognitive Communications and Networking}, 
  title={{MIMO} Radar Spectrally Compatible Waveform Design via Inequality Constrained Manifold Optimization}, 
  year={2025},
  volume={11},
  number={1},
  pages={362-374},
  doi={10.1109/TCCN.2024.3435364}}

@ARTICLE{7202844,
  author={Aubry, Augusto and De Maio, Antonio and Naghsh, Mohammad Mahdi},
  journal={IEEE Journal of Selected Topics in Signal Processing}, 
  title={Optimizing Radar Waveform and Doppler Filter Bank via Generalized Fractional Programming}, 
  year={2015},
  volume={9},
  number={8},
  pages={1387-1399},
  doi={10.1109/JSTSP.2015.2469259}}

@ARTICLE{8706639,
  author={Alaee-Kerahroodi, Mohammad and Modarres-Hashemi, Mahmoud and Naghsh, Mohammad Mahdi},
  journal={IEEE Transactions on Signal Processing}, 
  title={Designing Sets of Binary Sequences for {MIMO} Radar Systems}, 
  year={2019},
  volume={67},
  number={13},
  pages={3347-3360},
  doi={10.1109/TSP.2019.2914878}}

@ARTICLE{7967829,
  author={Kerahroodi, Mohammad Alaee and Aubry, Augusto and De Maio, Antonio and Naghsh, Mohammad Mahdi and Modarres-Hashemi, Mahmoud},
  journal={IEEE Transactions on Signal Processing}, 
  title={A Coordinate-Descent Framework to Design Low {PSL}/{ISL} Sequences}, 
  year={2017},
  volume={65},
  number={22},
  pages={5942-5956},
  doi={10.1109/TSP.2017.2723354}}

@ARTICLE{8903541,
  author={Sun, Yinghao and Fan, Huayu and Mao, Erke and Liu, Quanhua and Long, Teng},
  journal={IEEE Transactions on Aerospace and Electronic Systems}, 
  title={Range-Doppler Sidelobe Suppression for Pulse-Diverse Waveforms}, 
  year={2020},
  volume={56},
  number={4},
  pages={2835-2849},
  doi={10.1109/TAES.2019.2954152}}

@ARTICLE{9724170,
  author={Ye, Zhifan and Zhou, Zhengchun and Fan, Pingzhi and Liu, Zilong and Lei, Xianfu and Tang, Xiaohu},
  journal={IEEE Journal on Selected Areas in Communications}, 
  title={{Low Ambiguity Zone}: Theoretical Bounds and {Doppler-Resilient} Sequence Design in Integrated Sensing and Communication Systems}, 
  year={2022},
  volume={40},
  number={6},
  pages={1809-1822},
  doi={10.1109/JSAC.2022.3155510}}

@ARTICLE{7469294,
  author={Zhang, Jindong and Shi, Changli and Qiu, Xiaoyan and Wu, Yue},
  journal={IEEE Sensors Journal}, 
  title={Shaping Radar Ambiguity Function by  $L$ -Phase Unimodular Sequence}, 
  year={2016},
  volume={16},
  number={14},
  pages={5648-5659},
  doi={10.1109/JSEN.2016.2567643}}

@article{CHEN2023109075,
title = {On designing good doppler tolerance waveform with low {PSL} of ambiguity function},
journal = {Signal Processing},
volume = {210},
pages = {109075},
year = {2023},
issn = {0165-1684},
doi = {https://doi.org/10.1016/j.sigpro.2023.109075},
url = {https://www.sciencedirect.com/science/article/pii/S0165168423001494},
author = {Zihao Chen and Junli Liang and Keman Song and Yuqiao Yang and Xiaobo Deng}
}

@ARTICLE{9409634,
  author={Imani, Sadjad and Nayebi, Mohammad Mahdi},
  journal={IEEE Transactions on Aerospace and Electronic Systems}, 
  title={A Coordinate Descent Framework for Beampattern Design and Waveform Synthesis in {MIMO} Radars}, 
  year={2021},
  volume={57},
  number={6},
  pages={3552-3562},
  doi={10.1109/TAES.2021.3074207}}

@ARTICLE{9745120,
  author={Alaee-Kerahroodi, Mohammad and Raei, Ehsan and Kumar, Sumit and M. R. R., Bhavani Shankar},
  journal={IEEE Sensors Journal}, 
  title={Cognitive Radar Waveform Design and Prototype for Coexistence With Communications}, 
  year={2022},
  volume={22},
  number={10},
  pages={9787-9802},
  doi={10.1109/JSEN.2022.3163548}}

@ARTICLE{9440807,
  author={Raei, Ehsan and Alaee-Kerahroodi, Mohammad and Shankar, M.R. Bhavani},
  journal={IEEE Transactions on Signal Processing}, 
  title={Spatial- and Range- {ISLR} Trade-Off in {MIMO} Radar Via Waveform Correlation Optimization}, 
  year={2021},
  volume={69},
  number={},
  pages={3283-3298},
  doi={10.1109/TSP.2021.3082460}}

@ARTICLE{10679908,
  author={Chen, Yun and Chen, Zixuan and Zhang, Yunhua and Yang, Jiefang and Li, Dong},
  journal={IEEE Transactions on Signal Processing}, 
  title={Joint Design of Doppler Resilient Unimodular Discrete Phase Sequence Waveform and Receiving Filter for Multichannel Radar}, 
  year={2024},
  volume={72},
  number={},
  pages={4207-4221},
  doi={10.1109/TSP.2024.3458175}}

@ARTICLE{9257002,
  author={Hu, Jinfeng and Wei, Zhiyong and Li, Yuzhi and Li, Huiyong and Wu, Jie},
  journal={IEEE Transactions on Aerospace and Electronic Systems}, 
  title={Designing Unimodular Waveform(s) for {MIMO} Radar by Deep Learning Method}, 
  year={2021},
  volume={57},
  number={2},
  pages={1184-1196},
  doi={10.1109/TAES.2020.3037406}}

@ARTICLE{10453983,
  author={Qiu, Zizhou and Duan, Keqing and Wang, Yongliang and Liao, Zhipeng and He, Jinjun},
  journal={IEEE Transactions on Geoscience and Remote Sensing}, 
  title={Designing Constant Modulus Approximate Binary Phase Waveforms for Multitarget Detection in {MIMO} Radar Using {LSTM} Networks}, 
  year={2024},
  volume={62},
  number={},
  pages={1-20},
  doi={10.1109/TGRS.2024.3371532}}

@ARTICLE{11079603,
  author={Zhong, Kai and Jia, Hezhe and Wang, Ren and Hu, Jinfeng and Teh, Kah Chan and Liu, Jun and Wang, Yuankai and Li, Huiyong and Li, Chaohai and Yu, Xianxiang},
  journal={IEEE Transactions on Cognitive Communications and Networking}, 
  title={Design of Discrete-Phase Orthogonal Waveforms for Cognitive {MIMO} Systems}, 
  year={2025},
  volume={},
  number={},
  pages={1-1},
  doi={10.1109/TCCN.2025.3588733}}

@ARTICLE{9618146,
  author={Xu, Wangyang and Gan, Lu and Huang, Chongwen},
  journal={IEEE Transactions on Cognitive Communications and Networking}, 
  title={A Robust {Deep Learning-Based} Beamforming Design for {RIS-Assisted} Multiuser {MISO} Communications With Practical Constraints}, 
  year={2022},
  volume={8},
  number={2},
  pages={694-706},
  doi={10.1109/TCCN.2021.3128605}}

@ARTICLE{10380119,
  author={Zhong, Kai and Hu, Jinfeng and Zhao, Ziwei and Yu, Xianxiang and Cui, Guolong and Liao, Bin and Hu, Haotian},
  journal={IEEE Transactions on Aerospace and Electronic Systems}, 
  title={{MIMO} Radar Unimodular Waveform Design With Learned {Complex Circle Manifold} Network}, 
  year={2024},
  volume={60},
  number={2},
  pages={1798-1807},
  doi={10.1109/TAES.2023.3344391}}

@ARTICLE{9404363,
  author={Wang, Fulai and Pang, Chen and Zhou, Jian and Li, Yongzhen and Wang, Xuesong},
  journal={IEEE Geoscience and Remote Sensing Letters}, 
  title={Design of Complete Complementary Sequences for Ambiguity Functions Optimization With a PAR Constraint}, 
  year={2022},
  volume={19},
  number={},
  pages={1-5},
  doi={10.1109/LGRS.2021.3071249}}

@ARTICLE{9684877,
  author={Wang, Fulai and Yin, Jiapeng and Pang, Chen and Li, Yongzhen and Wang, Xuesong},
  journal={IEEE Transactions on Geoscience and Remote Sensing}, 
  title={A Unified Framework of Doppler Resilient Sequences Design for Simultaneous Polarimetric Radars}, 
  year={2022},
  volume={60},
  number={},
  pages={1-15},
  doi={10.1109/TGRS.2022.3144385}}

@ARTICLE{9093027,
  author={Huang, Wenjie and Naghsh, Mohammad Mahdi and Lin, Ronghao and Li, Jian},
  journal={IEEE Transactions on Aerospace and Electronic Systems}, 
  title={Doppler Sensitive Discrete-Phase Sequence Set Design for {MIMO} Radar}, 
  year={2020},
  volume={56},
  number={6},
  pages={4209-4223},
  doi={10.1109/TAES.2020.2988623}}

@ARTICLE{9765443,
  author={Wang, Fulai and Feng, Sijia and Yin, Jiapeng and Pang, Chen and Li, Yongzhen and Wang, Xuesong},
  journal={IEEE Transactions on Geoscience and Remote Sensing}, 
  title={Unimodular Sequence and Receiving Filter Design for Local Ambiguity Function Shaping}, 
  year={2022},
  volume={60},
  number={},
  pages={1-12},
  doi={10.1109/TGRS.2022.3171253}}

@ARTICLE{9628070,
  author={Zhou, Kai and Quan, Sinong and Li, Dexin and Liu, Tao and He, Feng and Su, Yi},
  journal={IEEE Transactions on Geoscience and Remote Sensing}, 
  title={Waveform and Filter Joint Design Method for Pulse Compression Sidelobe Reduction}, 
  year={2022},
  volume={60},
  number={},
  pages={1-15},
  doi={10.1109/TGRS.2021.3131590}}

@ARTICLE{10466432,
  author={Chen, Yutao and Cheng, Yuanbo and Lin, Ronghao and So, Hing Cheung and Li, Jian},
  journal={IEEE Transactions on Signal Processing}, 
  title={Joint Design of Binary Probing Sequence Sets and Receive Filter Banks for {MIMO} {PM{CW}} Radar}, 
  year={2024},
  volume={72},
  number={},
  pages={1620-1633},
  doi={10.1109/TSP.2024.3375646}}

@Article{rs15153877,
AUTHOR = {Chen, Yun and Zhang, Yunhua and Li, Dong and Yang, Jiefang},
TITLE = {Joint Design of Complementary Sequence and Receiving Filter with High Doppler Tolerance for Simultaneously Polarimetric Radar},
JOURNAL = {Remote Sensing},
VOLUME = {15},
YEAR = {2023},
NUMBER = {15},
ARTICLE-NUMBER = {3877},
URL = {https://www.mdpi.com/2072-4292/15/15/3877},
ISSN = {2072-4292},
DOI = {10.3390/rs15153877}
}

@ARTICLE{9018020,
  author={Wu, Zhong-Jie and Zhou, Zhi-Quan and Wang, Chen-Xu and Li, Ying-Chun and Zhao, Zhan-Feng},
  journal={IEEE Sensors Journal}, 
  title={Doppler Resilient Complementary Waveform Design for Active Sensing}, 
  year={2020},
  volume={20},
  number={17},
  pages={9963-9976},
  doi={10.1109/JSEN.2020.2976525}}

@ARTICLE{10003252,
  author={Wang, Fulai and Xia, Xiang-Gen and Pang, Chen and Cheng, Xu and Li, Yongzhen and Wang, Xuesong},
  journal={IEEE Transactions on Geoscience and Remote Sensing}, 
  title={Joint Design Methods of Unimodular Sequences and Receiving Filters With Good Correlation Properties and Doppler Tolerance}, 
  year={2023},
  volume={61},
  number={},
  pages={1-14},
  doi={10.1109/TGRS.2022.3233094}}

@ARTICLE{8239836,
  author={Wu, Linlong and Babu, Prabhu and Palomar, Daniel P.},
  journal={IEEE Transactions on Signal Processing}, 
  title={Transmit Waveform/Receive Filter Design for {MIMO} Radar With Multiple Waveform Constraints}, 
  year={2018},
  volume={66},
  number={6},
  pages={1526-1540},
  doi={10.1109/TSP.2017.2787115}}

@ARTICLE{7728070,
  author={Beauchamp, Robert M. and Tanelli, Simone and Peral, Eva and Chandrasekar, V.},
  journal={IEEE Transactions on Geoscience and Remote Sensing}, 
  title={Pulse Compression Waveform and Filter Optimization for Spaceborne Cloud and Precipitation Radar}, 
  year={2017},
  volume={55},
  number={2},
  pages={915-931},
  doi={10.1109/TGRS.2016.2616898}}

@ARTICLE{8844976,
  author={Esmaeili-Najafabadi, Hamid and Leung, Henry and Moo, Peter W.},
  journal={IEEE Transactions on Aerospace and Electronic Systems}, 
  title={Unimodular Waveform Design With Desired Ambiguity Function for Cognitive Radar}, 
  year={2020},
  volume={56},
  number={3},
  pages={2489-2496},
  doi={10.1109/TAES.2019.2942411}}

@ARTICLE{9987663,
  author={Chen, Yutao and Lin, Ronghao and Cheng, Yuanbo and Li, Jian},
  journal={IEEE Transactions on Signal Processing}, 
  title={Joint Design of Periodic Binary Probing Sequences and Receive Filters for {PM{CW}} Radar}, 
  year={2022},
  volume={70},
  number={},
  pages={5996-6010},
  doi={10.1109/TSP.2022.3229629}}

@article{RAEI2023108914,
title = {Generalized waveform design for sidelobe reduction in {MIMO} radar systems},
journal = {Signal Processing},
volume = {206},
pages = {108914},
year = {2023},
issn = {0165-1684},
doi = {https://doi.org/10.1016/j.sigpro.2022.108914},
url = {https://www.sciencedirect.com/science/article/pii/S0165168422004534},
author = {Ehsan Raei and Mohammad Alaee-Kerahroodi and Prabhu Babu and M.R. {Bhavani Shankar}}
}

@article{https://doi.org/10.1049/rsn2.12192,
author = {Huang, Zhongrui and Shi, Yingchun and Tang, Bo and Zhang, Junning},
title = {Unimodular multiple-input-multiple-output radar wave-form design with desired correlation properties},
journal = {IET Radar, Sonar \& Navigation},
volume = {16},
number = {3},
pages = {412-425},
doi = {https://doi.org/10.1049/rsn2.12192},
url = {https://ietresearch.onlinelibrary.wiley.com/doi/abs/10.1049/rsn2.12192},
eprint = {https://ietresearch.onlinelibrary.wiley.com/doi/pdf/10.1049/rsn2.12192},
year = {2022}
}

@ARTICLE{8979323,
  author={Feraidooni, Mohamad Mahdi and Gharavian, Davood and Alaee-Kerahroodi, Mohammad and Imani, Sadjad},
  journal={IEEE Communications Letters}, 
  title={A Coordinate Descent Framework for Probing Signal Design in Cognitive {MIMO} Radars}, 
  year={2020},
  volume={24},
  number={5},
  pages={1115-1118},
  doi={10.1109/LCOMM.2020.2971210}}

@ARTICLE{8890867,
  author={Yu, Xianxiang and Cui, Guolong and Yang, Jing and Kong, Lingjiang},
  journal={IEEE Transactions on Vehicular Technology}, 
  title={{MIMO} Radar Transmit–Receive Design for Moving Target Detection in Signal-Dependent Clutter}, 
  year={2020},
  volume={69},
  number={1},
  pages={522-536},
  doi={10.1109/TVT.2019.2951399}}

@ARTICLE{11168247,
  author={Wang, Wei and Li, Chunshen and Zeng, Bixin and Chen, Lieke and Sun, Liang and Luo, Kai and Chen, Da},
  journal={IEEE Transactions on Mobile Computing}, 
  title={Mitigating Interference for Automotive Millimeter-Wave Radar Perception in Dense Traffic Scenarios}, 
  year={2026},
  volume={25},
  number={2},
  pages={2076-2090},
  doi={10.1109/TMC.2025.3610963}}

@ARTICLE{11408926,
  author={Yang, Pengfei and Liu, Minyang and Zhong, Ting and Zhou, Fan and Zhang, Kunpeng and Yu, Philip S.},
  journal={IEEE Transactions on Mobile Computing}, 
  title={Interpolate-Then-Detect: A New Framework of {FM{CW}} Radar Object Detection With Frame Interpolation for Autonomous Driving}, 
  year={2026},
  volume={},
  number={},
  pages={1-16},
  doi={10.1109/TMC.2026.3667477}}

@ARTICLE{11457727,
  author={Deng, Kaikai and Zhao, Dong and Wang, Shuyue and Zheng, Wenxin and Zhang, Zihan and Ma, Huadong},
  journal={IEEE Transactions on Mobile Computing}, 
  title={Argus: {3D} Radar Point Cloud Registration Using Complementary Cameras for Infrastructure-Assisted Autonomous Driving}, 
  year={2026},
  volume={},
  number={},
  pages={1-18},
  doi={10.1109/TMC.2026.3678325}}

@ARTICLE{11184442,
  author={Lodhi, Shikhar Singh and Kumar, Neetesh and Pandey, Pradumn Kumar},
  journal={IEEE Transactions on Mobile Computing}, 
  title={Semi-Markov Options Enabled DDPG Method for Autonomous Vehicle Overtaking With LiDAR and RADAR Fusion}, 
  year={2026},
  volume={25},
  number={3},
  pages={3247-3260},
  doi={10.1109/TMC.2025.3615629}}

\begin{IEEEbiography}[{\includegraphics[width=1in,height=1.25in,clip,keepaspectratio]{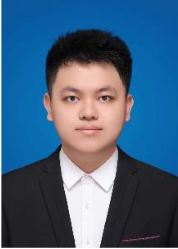}}]{Junpeng Ma}
received the B.E. degree in Computer Science and Technology from North China Electric Power University, Baoding, China, in 2025. He is currently pursuing the graduate degree with the School of Software, Northwestern Polytechnical University, Xi’an, China. His research interests include intelligent transportation systems, integrated Sensing and Communication (ISAC) and signal processing.
\end{IEEEbiography}

\begin{IEEEbiography}[{\includegraphics[width=1in,height=1.25in,clip,keepaspectratio]{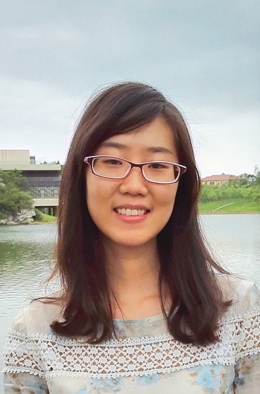}}]{Yuke Li (Member, IEEE)}
received the B.E. degree in Communication Engineering from Xidian University, Xi’an, China, in 2014, and the Ph.D. degree in Control Theory and Control Engineering from the State Key Laboratory of Management and Control for Complex Systems, Institute of Automation, Chinese Academy of Sciences, Beijing, China, in 2019. From October 2021 to October 2023, she was a postdoctoral research fellow with Huawei Technologies Co., Ltd. She is currently an Associate Professor with the School of Software, Northwestern Polytechnical University. Her research interests include intelligent transportation systems, signal processing, integrated Sensing and Communication (ISAC), and the Internet of Things.
\end{IEEEbiography}

\begin{IEEEbiography}[{\includegraphics[width=1in,height=1.25in,clip,keepaspectratio]{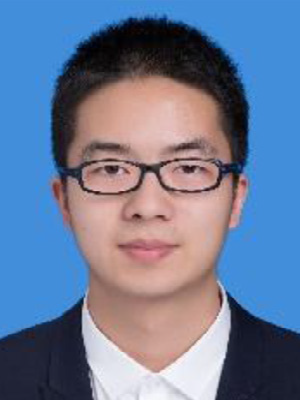}}]{Junbo Wang (Member, IEEE)}
received the B.S. degree from software engineering of Northeastern University in 2014, the Ph.D. degree from the National Laboratory of Pattern Recognition (NLPR), Institute of Automation, Chinese Academy of Sciences (CASIA), Beijing, China, in 2020. Now he is the associate professor at the School of Software, Northwestern Polytechnical University. His research interests include video summarization, video captioning, image captioning and deep learning.
\end{IEEEbiography}

\begin{IEEEbiography}[{\includegraphics[width=1in,height=1.25in,clip,keepaspectratio]{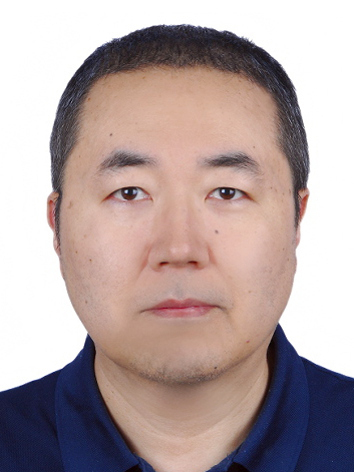}}]{Yongxing Zhou (Fellow, IEEE)}
received his Ph.D degree in Tsinghua University, China in 2002. He is now a professor with Beijing University of Posts and Telecommunications (BUPT). He has been a prestigious wireless telecom technical leader in the development of MIMO and smart spectrum access innovative technologies for 4G and 5G cellular communication standards and products. Before joining BUPT in September 2025, he was with Huawei as Principal Scientist of Standard $\&$ Patent and Huawei Device communication standard Principal Expert. He has led 16 years of Huawei 4G/5G/Device Communication research and standardization including multiple paradigm shifts such as Linear Combination Double MIMO Codebook, Flexible Bandwidth Part (a.k.a. BWP), beam based cellular initial access and the world’s first commercial satellite-smart phone direct communication protocols that have shaped landscape of 5G mobile communications.
Dr. Zhou’s current research interest includes 5G Advanced and 6G technologies such as Artificial Intelligence (AI) Communication, Satellite Communication, Integrated Sensing and Communication (ISAC), cell free MIMO and reconfigurable intelligent surface (RIS).
Dr. Zhou has more than 200 Granted U.S. Utility Patents and more than 100 Standard Essential Patents (SEP) have been widely used in global commercial 4G/5G base stations and terminals.
\end{IEEEbiography}


\end{document}